\documentclass[letterpaper,onecolumn,floatfix,nobibnotes,nofootinbib]{revtex4}
\usepackage{graphicx}

\begin{document}

\title{Discreteness-Induced Transitions in Autocatalytic Systems}
\author{Yuichi Togashi}
\email{togashi@complex.c.u-tokyo.ac.jp}
\author{Kunihiko Kaneko}
\affiliation{Department of Basic Science, School of Arts and Sciences,
The University of Tokyo,
Komaba, Meguro, Tokyo 153-8902, Japan}
\date{July 29, 2004}

\maketitle
\section*{Abstract}
To study the dynamics of chemical processes, we often adopt rate equations
to observe the change in chemical concentrations.
However, when the number of the molecules is small,
the fluctuations cannot be neglected.
We often study the effects of fluctuations with the help of stochastic
differential equations.

Chemicals are composed of molecules on a microscopic level.
In principle, the number of molecules must be an integer,
which must only change discretely.
However, in analysis using stochastic differential equations,
the fluctuations are regarded as continuous changes.
This approximation can only be valid if applied to fluctuations
that involve a sufficiently large number of molecules.
In the case of extremely rare chemical species, the actual discreteness
of the molecules may critically affect the dynamics of the system. 

To elucidate the effects of the discreteness, we study an autocatalytic
system consisting of several interacting chemical species with a
small number of molecules through stochastic particle simulations.
We found novel states, which were characterized as an extinction of molecule
species, due to the discrete nature of the molecules.
We also observed a strong dependence of the chemical concentrations
on the size of the system, which was caused by transitions
to the novel states. 

\section{Introduction: Discrete Reaction Systems}

In nature, there exists a variety of systems that involve
chemical reactions.
Some are on a geographical-scale, while others on a nano-scale.
Chemical reactions are an integral part of life,
including all living forms of life. 

To study the dynamics of reaction systems, we often adopt rate equations
in order to observe the change in chemical concentrations.
In rate equations, we regard the concentrations as continuous variables;
the rate of the reaction as a function of the concentrations.
In macroscopic systems, there are a vast number of molecules;
thus, continuous representations are usually applicable. 

When the concentration of a certain chemical is small,
fluctuations in the reactions or flow can be significant.
We often handle such systems with the help of stochastic differential equations,
in which we regard noise as a continuum description of the fluctuations
\cite{Nicolis,Kampen}.
Such an approximation is useful when the number of molecules is
intermediate.
The employment of stochastic differential equations led to some important discoveries
such as noise-induced order \cite{NIO},
noise-induced phase transitions \cite{NIP},
and stochastic resonance \cite{SR}.

In stochastic differential equations, still quantities of chemicals are
regarded as continuous variables.
Essentially, on a microscopic level, chemicals are composed of molecules.
The number of molecules should be an integer ($0$, $1$, $2$, $\cdots$),
which changes discretely.
Fluctuations are derivatives of discrete stochastic changes;
thus, continuum descriptions of fluctuations are not always
appropriate and can be doubted.
For chemicals with a small number of molecules of the order of $1$,
a single molecule is extremely significant;
therefore, the discreteness in the number is significant.

Biological cells appear to be a good example.
The size of the cells is of the order of microns,
in which nano-scale ``quantum'' effects can be ignored.
However, in cells, some chemicals act
at extremely low concentrations of the order of pM or nM.
Assuming that the typical volume of a cell ranges from $1$ to
$10^{3}$ $\mu m^{3}$, the concentration of
one molecule in the cell volume
corresponds to $1.7$ pM--$1.7$ nM.
It is probable that the molecular numbers of some chemicals in a cell are
of the order of $1$, or sometimes reach $0$.

If such chemicals play only a minor role,
we can safely ignore these chemicals.
However, this is not always the case.
In biological systems, chemical species with a small number of
molecules may critically affect the behavior of the entire system.
For example, there exist only one or a few copies of genetic molecules
such as DNA, which are important to characterize the behavior,
in each cell.
Further, some experiments show that doses of particular chemicals
at concentrations of the order of pM or fM may alter the behavior of the cells
(e.g., \cite{TGF1,IL1B}).
Biological systems also include positive-feedback mechanisms
such as autocatalytic reactions,
which may amplify single molecular changes to a macroscopic level.
The effects due to small molecular numbers in cells
have been noticed only recently,
both theoretically \cite{McAdams,Arkin1} and experimentally \cite{Elowitz2002}.

At present, we focus on the possible effects of molecular discreteness.
To study such effects, we should adopt an appropriate method
to handle molecular discreteness.
Some numerical methods to investigate reaction systems
that take into account discreteness and
stochasticity already exist (we briefly review these methods; see Appendix).
Among the methods, we adopted Gillespie's direct method,
which is popular and frequently used.

Furthermore, some works related to molecular discreteness also exist.
For example,
Blumenfeld et al. showed that the mass action law may breakdown in a small system
\cite{Blumenfeld1991,Blumenfeld}.
Stange et al. studied the synchronization of the turnover cycle of enzymes
\cite{Mikhailov2C,Mikhailov2002}.

We regard it important to identify the phenomena for which
molecular discreteness is essential.

Through stochastic simulations,
we show that discreteness can induce transitions to novel states in
autocatalytic systems \cite{Togashi},
which may affect macroscopic chemical concentrations \cite{Togashi2}.

\section{Discreteness-Induced Transitions}\label{sect:tr1}

\subsection{Model}

We consider a simple autocatalytic network (loop) with $k$ chemicals.
We consider $X_{i}$ chemicals and assume
\[
X_{i} + X_{i+1} \rightarrow 2X_{i+1}
\]
reactions between these chemicals
($i=1$, $2$, $\cdots$ , $k$; $X_{k+1} \equiv X_{1}$).
All the reactions are irreversible.

For the reactor, we assume a well-stirred container with volume $V$.
The set of $N_{i}$, the number of $X_{i}$ molecules, determines the state
of the system.
The container is in contact with a chemical reservoir, in which
the concentration of $X_{i}$ is fixed at $s_{i}$.
The flow rate of $X_{i}$ between the container and the reservoir is $D_{i}$,
which corresponds to the probability of the flowing out of a molecule
per time unit\footnote{$D_{i}$ is the diffusion rate across the surface
of the container.
Here, we choose the flow proportional to $V$,
to have a well-defined continuum limit.
One might assume the flow proportional to $V^{2/3}$,
considering the area of the surface.
By rescaling $D$, the model can be rewritten into the case
with $V^{2/3}$ for finite $V$.}.

We can consider the continuum limit as $V \rightarrow \infty$.
In the continuum limit, the change of $x_{i}$,
the chemical concentration $X_{i}$ in the container,
follows the rate equation
\begin{equation}
\frac{dx_{i}}{dt} = r_{i-1}x_{i-1}x_{i} - r_{i}x_{i}x_{i+1} + D_{i}(s_{i} - x_{i}),
\label{eqn:tr1-rate}
\end{equation}
where $r_{i}$ is the rate constant of
the reaction $X_{i} + X_{i+1} \rightarrow 2X_{i+1}$, and $X_{0} \equiv X_{k}$.

For simplicity, we consider the case with equivalent chemical species,
given as $r_{i}=r$, $D_{i}=D$, and $s_{i}=s$ for all $i$ ($r$, $D$, $s > 0$).
By this assumption, the rate equation has only one attractor:
a stable fixed point $x_{i} = s$ for all $i$.
For any initial condition, each $x_{i}$ converges to $s$, the fixed point value.
Around the fixed point, $x_{i}$ vibrates
with the frequency $\omega_{p} \equiv rs / \pi$.

If the number of molecules is finite but fairly large,
we can estimate the dynamical behavior of the system
using a Langevin equation, obtained by adding a noise term to the rate equation.
Each concentration $x_{i}$ fluctuates and vibrates around the fixed point.
An increase in the noise (corresponding to a decrease in the number of molecules)
merely boosts the fluctuation.

However, when the number of molecules is small,
the behavior of the system is completely different.
First, we investigate the case when $k=4$,
which is the smallest number of species
to show the novel states described below.

\subsection{Novel States Induced by the Discreteness}

Subsequently, we investigate the dynamical behavior of the system
with a small number of molecules.
In order to detect the phenomena for which the discreteness
of the number of molecules
is crucial, we employ stochastic simulations.

Here, we adopt Gillespie's direct method\footnote{For systems with a large number of chemical species, the next reaction method
would be more suitable.}.
The frequency (expected number per unit time)
\begin{itemize}
\item of the reaction $X_{i} + X_{i+1} \rightarrow 2X_{i+1}$ is
$P_{Ri} \equiv rx_{i}x_{i+1}V = rN_{i}N_{i+1}/V$;
\item of the outflow of $X_{i}$ is $P_{Oi} \equiv Dx_{i}V = DN_{i}$;
\item of the inflow of $X_{i}$ is $P_{Ii} \equiv DsV$.
\end{itemize}
In the continuum limit ($V \rightarrow \infty$), these frequencies
agree with the rate equation.
We calculate these frequencies with the current $N_{i}$,
and stochastically decide when and which event will occur next.

In this case, by an appropriate conversion of $D$, $V$, and $t$,
we can set $r$ and $s$ to be $1$ without loss of the generality
($rs/D$ and $sV$ are the only independent parameters).
We assume that $r=1$ and $s=1$ for the purpose of further discussion.
The total number of molecules in the container,
$N_{tot} = \sum N_{i}$, is approximately $4sV$ on an average.
By varying $V$, we can control the average number of molecules
without changing the continuum limit.

First, we consider the case of a large $V$, i.e.,
both the number of molecules in the container
and flow of molecules between the container and the reservoir are large.
As expected, the behavior of the system is similar to
that of the rate equation with noise.
As shown in Fig. \ref{fig:A-512}, each $N_{i}$ fluctuates and vibrates around the
fixed point value $N_{i}=V$ (i.e. $x_{i}=1$).
This is still in the realm of stochastic differential equations.

However, when $V$ is small, we observe novel states
that do not exist in the continuum limit.
As shown in Fig. \ref{fig:A-032},
continuous vibrations disappear.
Furthermore, two chemicals are dominant
and the other two are mostly extinct ($N_{i}=0$).
In Fig. \ref{fig:A-032}, at $t < 520$, $N_{1}$ and $N_{3}$ dominate the system and
$N_{2} = N_{4} = 0$ for the most part.
We call such a state the 1-3 rich state.
Reversely, at $t > 520$, $N_{2}$ and $N_{4}$ are large and
usually $N_{1} = N_{3} = 0$.
We call this state the 2-4 rich state.

These states appear because of the following reason.
In this system, the production of $X_{i}$ molecules requires at least
one $X_{i}$ molecule as a catalyst.
If $X_{i}$ becomes extinct, the production of $X_{i}$ halts.
$N_{i}$ never regains before an $X_{i}$ molecule flows in.

In the rate equation (eq. (\ref{eqn:tr1-rate})),
the concentration $x_{i}$ is a continuous variable,
which can be an infinitesimal but positive value.
The consumption rate of $x_{i}$ is proportional to $x_{i}$ itself;
thus, $x_{i}$ cannot reach $0$ exactly within finite time,
even if it can go to $0$ asymptotically as $t \rightarrow \infty$.

In fact, the number of molecules must be an integer.
Transitions from $N_{i}=1$ to $N_{i}=0$ are probabilistic
and may happen in finite time.
For transitions to occur, it is important that
the consumption rate of $x_{i}$ does not converge to $0$ at $N_{i} = 1$.
The average interval of a molecule flowing in is $1 / DV$ for
each chemical.
If $D$ and $V$ are small enough to ensure that
the inflow interval is longer than the time scale of the reactions,
it is likely that the state of the system $N_{i}=1$ drops to $N_{i}=0$
before an $X_{i}$ molecule enters.

When $N_{i}$ reaches $0$, $N_{i+2}$ is also likely to become $0$.
For example, if we assume that $N_{2} = 0$,
then $N_{1}$ is likely to increase because the consumption of $X_{1}$ halts;
$N_{3}$ is likely to decrease because the production of $X_{3}$ halts.
Thus, this results in $N_{1} > N_{3}$.
When $N_{1} > N_{3}$, the consumption rate of $X_{4}$ is larger
than the production rate of $X_{4}$;
therefore, $N_{4}$ starts to decrease and often reaches $0$.
When $N_{2} = N_{4} = 0$, all the reactions stop.
The system stays at $N_{2} = N_{4} = 0$ for a long time as compared with
the ordinary time scale of the reactions ($\sim 1/r$).
This is the 1-3 rich state.

In the 1-3 rich state, the system alternately switches
between $N_{1} > N_{3}$ and $N_{1} < N_{3}$.
We consider that the system is in the 1-3 rich state with $N_{1} > N_{3}$.
One $X_{2}$ molecule flowing in may resume
the reactions $X_{1} + X_{2} \rightarrow 2X_{2}$ and
$X_{2} + X_{3} \rightarrow 2X_{3}$.
Generally, the former is faster because $N_{1} > N_{3}$;
hence, $N_{2}$ is likely to increase.
Since $N_{4}$ = 0, the reactions are one-way;
$N_{1}$ decreases and $N_{3}$ increases.
When $N_{1} < N_{3}$, $N_{2}$ starts to decrease.
Finally, when $N_{2}$ returns to $0$, the reactions halt again.
The system stays in the 1-3 rich state, until $N_{1}$ reaches $0$.
In the same manner, the inflow of $X_{4}$ can switch the system
from $N_{1} < N_{3}$ to $N_{1} > N_{3}$ in the 1-3 rich state.
Consequently, we observe successive switching between
$N_{1} > N_{3}$ and $N_{1} < N_{3}$.
In the 2-4 rich state, the system switches
between $N_{2} > N_{4}$ and $N_{2} < N_{4}$.

In this manner, even one molecule can switch the system
within the 1-3 or 2-4 rich states.
We name these states ``switching states''.

Now, we investigate some properties of the switching states.
We introduce an index $z \equiv (x_{1} + x_{3}) - (x_{2} + x_{4})$ as
a characteristic of the switching states.
Around the fixed point of the rate equation, $z \approx 0$;
in the 1-3 rich state, $z \approx 4$;
in the 2-4 rich state, $z \approx -4$.

\begin{figure}
\begin{center}
\includegraphics[width=69mm]{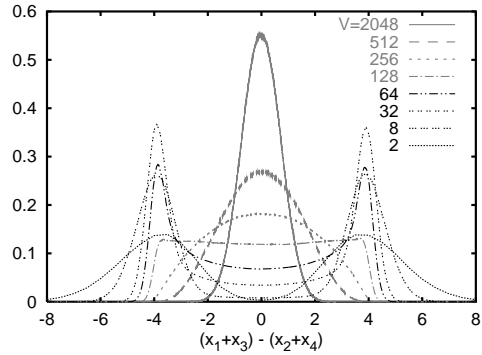}
\end{center}
\caption{The probability distribution of the index
$z = (x_{1}+x_{3}) - (x_{2}+x_{4})$,
sampled over $5\times 10^{6}$ time units.
($z$ is actually a discrete value.
Here, we show the distribution as a line graph for visibility.) $D=1/128$.
When $V$ is large ($V\ge 256$), there appears a single peak around $z=0$,
corresponding to the fixed point state $\forall i : x_{i}=1$.
For $V\le 64$, the distribution has double peaks around $z=\pm 4$.
The peak $z \approx 4$ corresponds to the 1-3 rich state,
with $N_{1}+N_{3}\approx 4V$, $N_{2}=N_{4}=0$.
$z \approx -4$ corresponds to the 2-4 rich state as well.}
\label{fig:A-ac-bd}
\end{figure}

The distribution of $z$ is shown in Fig. \ref{fig:A-ac-bd}.
When $V \ge 128$, a single peak appears around $z=0$,
which corresponds to the fixed point.
By decreasing $V$, the peak broadens with fluctuations.
When $V \le 32$, double peaks appear at $z \approx \pm 4$,
which correspond to the switching states.
We clearly observe a symmetry-breaking transition between
a continuous vibration around the fixed point with large $V$
and the switching states with small $V$.
This is a discreteness-induced transition (DIT)
that occurs with decrease of $V$,
which is not seen in continuum descriptions.

\begin{figure}
\begin{center}
\includegraphics[width=69mm]{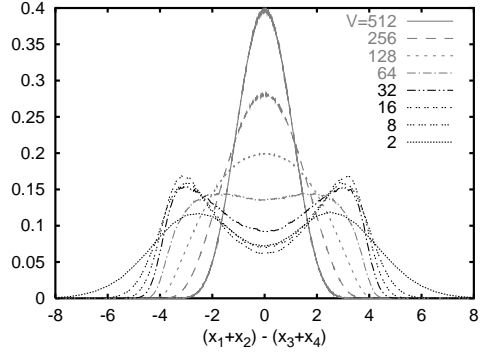}
\end{center}
\caption{The probability distribution of the index
$y = (x_{1}+x_{2}) - (x_{3}+x_{4})$,
sampled over $5\times 10^{6}$ time units.
$D=1/128$.
When $V$ is large, there appears a single peak around $y=0$,
corresponding to the fixed point.
When $V$ is small and the system is in the switching states,
the index $y$ shows an imbalance between the rich (non-zero) chemicals.
For example, in the 1-3 rich state, $y$ corresponds to
$x_{1}-x_{3}$ since $N_{2}=N_{4}=0$ for the most part.
The distribution shows double peaks around $y=\pm 3$.
Assuming that $x_{1}+x_{3}=4$ and $x_{2}=x_{4}=0$ in the 1-3 rich state,
$y=\pm 3$ correspond to $(x_{1}, x_{3}) = (3.5, 0.5)$, $(0.5, 3.5)$.
Both the chemicals are likely to have a large imbalance.}
\label{fig:A-ab-cd}
\end{figure}

We introduce another index
$y \equiv (x_{1} + x_{2}) - (x_{3} + x_{4})$,
which represents the difference in concentrations:
$x_{1} - x_{3}$ in the 1-3 rich state
and $x_{2} - x_{4}$ in the 2-4 rich state.
The distribution of $y$ is shown in Fig. \ref{fig:A-ab-cd}.
There are double peaks around $y = \pm 3$,
which imply large imbalances, such as
$(N_{1}, N_{3}) = (3.5V, 0.5V)$ and $(0.5V, 3.5V)$, between
$N_{1}$ and $N_{3}$ in the 1-3 rich state
(as well as the 2-4 rich state).

\subsection{Single Molecular Switch}

We investigate some properties of the switching states.

First, we examine how each $N_{i}$ changes in a switching event.
We assume the 1-3 rich state with $N_{1}=N_{1 ini}$,
$N_{3}=N_{3 ini}$, and $N_{2}=N_{4}=0$.
Here, one $X_{2}$ molecule flows in ($N_{2}=1$) at $t=0$,
which starts up the reactions.
Assuming that $DV$ is so small that no more molecules flow in or out
throughout the switching, the total number of molecules
$N$ is conserved at $N = N_{1 ini} + N_{3 ini} + 1$,
and $N_{4}$ is always $0$.
We can represent the state with two variables, $N_{1}$ and $N_{2}$.

In this system, only the following types of reactions
\begin{itemize}
\item $X_{1} + X_{2} \rightarrow 2 X_{2}$ and
\item $X_{2} + X_{3} \rightarrow 2 X_{3}$
\end{itemize}
may change $N_{i}$; the others never take place.
$N_{1}$ monotonously decreases.
When $N_{2}$ reaches $0$, these reactions halt completely,
and the switching is completed.
Evidently, $N_{1 fin} + N_{3 fin} = N_{1 ini} + N_{3 ini} + 1$,
where $N_{1 fin}$ and $N_{3 fin}$ are the final values
of $N_{1}$ and $N_{3}$, respectively.

The frequency of the reaction $X_{1} + X_{2} \rightarrow 2 X_{2}$ is
\[
f_{1}(N_{1}, N_{2}) = N_{1} N_{2} / V,
\]
and that of $X_{2} + X_{3} \rightarrow 2 X_{3}$ is
\[
f_{2}(N_{1}, N_{2}) = N_{2} N_{3} / V = N_{2} (N - N_{1} - N_{2}) / V.
\]
We obtain the Master equation
\begin{eqnarray}
\frac{dP(N_{1}, N_{2}, t)}{dt} &=& f_{1}(N_{1}+1, N_{2}-1) P(N_{1}+1, N_{2}-1, t) + f_{2}(N_{1}, N_{2}+1) P(N_{1}, N_{2}+1, t)\nonumber\\
 & & {} - \left\{ f_{1}(N_{1}, N_{2}) + f_{2}(N_{1}, N_{2}) \right\} P(N_{1}, N_{2}, t)\nonumber\\
 &=& \frac{1}{V} \Bigl\{ (N_{1}+1) (N_{2}-1) P(N_{1}+1, N_{2}-1, t)\nonumber\\
 & & \quad {} + (N_{2}+1) (N - N_{1} - N_{2} - 1) P(N_{1}, N_{2}+1, t)\nonumber\\
 & & \quad {} - N_{2} (N - N_{2}) P(N_{1}, N_{2}, t) \Bigr\} \label{eqn:tr1-master}
\end{eqnarray}
for $P(N_{1}, N_{2}, t)$, the probability of residence
in the state $(N_{1}, N_{2})$ at time $t$.
The initial condition is $P(N_{1 ini}, 1, 0) = 1$;
otherwise $P(N_{1}, N_{2}, 0) = 0$.

We can easily follow the Master equation numerically.
We investigate the relationship between
the initial state $(N_{1 ini}, N_{3 ini})$ and
the final state $(N_{1 fin}, N_{3 fin})$.

If the first reaction is $X_{2} + X_{3} \rightarrow 2 X_{3}$,
$N_{2}$ instantaneously reaches $0$,
and as a result $N_{3 fin} = N_{3 ini} + 1$.
The system fails to switch.

\begin{figure}
\begin{center}
\includegraphics[width=69mm]{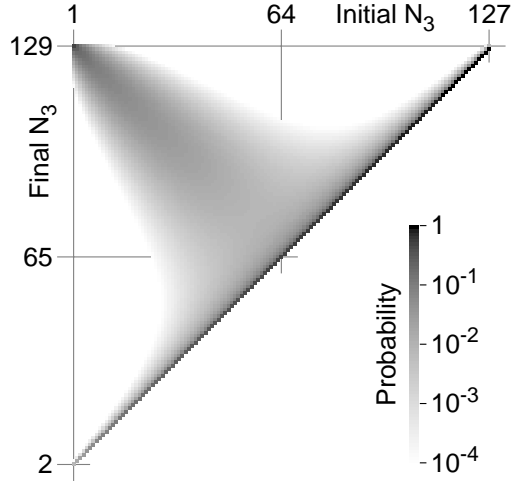}
\end{center}
\caption{Probability density for the switch from
$(N_{1}, N_{3}) = (N_{1 ini}, N_{3 ini})$ to $(N_{1 fin}, N_{3 fin})$.
We assume a switching event triggered by a single $X_{2}$ molecule
in the 1-3 rich state, and we set the initial conditions as $N_{2} = 1$ and $N_{4} = 0$.
Assuming that there is no further flow ($D=0$),
we numerically follow the master equation (eq. (\ref{eqn:tr1-master}))
until $t=50$ (sufficiently large to ensure $N_{2}=0$ for most cases).
For each initial condition, we show the transition probabilities
of the final states $(N_{1 fin}, N_{3 fin})$.
$V=32$, $N_{1 ini}+N_{3 ini}=128$.
The system shows high probabilities around $N_{3 fin}=N_{3 ini}+1$ (immediately terminated)
and $N_{3 fin}= N_{1 ini}$ (switching; $N_{1 ini} > N_{3 ini}$).}
\label{fig:A-swprob-128}
\end{figure}

If $N_{1 ini} > N_{3 ini}$, it is more probable that
the first reaction is $X_{1} + X_{2} \rightarrow 2 X_{2}$.
Subsequently, the system carries out further reactions.
In this case, it is probable that $N_{1 ini} \approx N_{3 fin}$,
i.e., the system swaps $N_{1}$ and $N_{3}$, as shown in Fig. \ref{fig:A-swprob-128}.
Consequently, we observe successive switching (as seen in Fig. \ref{fig:A-032}).

When $N_{1 ini} \gg N_{3 ini}$, the system is likely to
reach $N_{1 fin} = 0$ and break the 1-3 rich state.
A large imbalance between $N_{1}$ and $N_{3}$ results in
an unstable 1-3 rich state.


\subsection{Conditions for Switching States}\label{subsect:tr1-cond}

\begin{figure}
\begin{center}
\includegraphics[width=69mm]{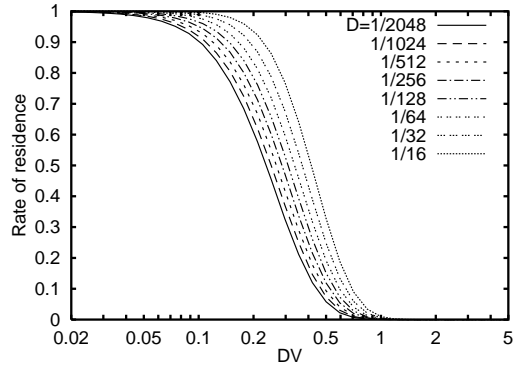}
\end{center}
\caption{The rate of residence of the switching states
plotted against $DV$, the inflow frequency,
sampled over $10^{6}$ ($V>1024$), $10^{7}$ ($32<V\le 1024$),
and $10^8$ ($V\le 32$) time units.
Here, we define the 1-3 rich state as continuation of the state
in which at least one of $N_{2}$ or $N_{4}$ is $0$ for $8$ time units or longer.
Thus, the system may contain states with only one or no chemical,
especially for small $V$.
We observe the switching states for $DV<1$.}
\label{fig:A-type1a}
\end{figure}

Now, we investigate the requirements for the transitions to
the switching states.
The rate of residence of the switching states
for several $D$ and $V$ is shown in Fig. \ref{fig:A-type1a}.
For approximately $DV < 1$, we observe the switching states.
If $DV < 0.1$, the system mostly stays in the switching states.

\begin{figure}
\begin{center}
\includegraphics[width=69mm]{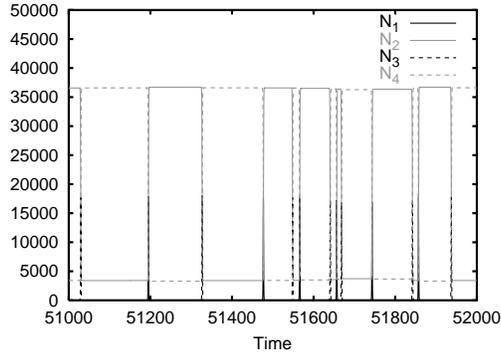}
\end{center}
\caption{Time series of $N_{i}$ for $V=10^{4}$ and $D=10^{-6}$.
For such a small $D$, we observe the switching states for relatively
large $V$.
In this case also, a single molecule can induce switching.
One $X_{1}$ molecule flows in ($N_{1}=1$), propagates to more than $10^{4}$,
and then returns to $0$.}
\label{fig:A-10000}
\end{figure}

Subsequently, we expect that switching states appear
even for large $V$ if $D$ is very small.
In fact, we observe switching states for $V=10^{4}$
as shown in Fig. \ref{fig:A-10000}.
Furthermore, in this case also,
a single molecule can induce switching.

Strictly speaking, if $DV$ is the same, the rate of residence is
a little smaller for larger $V$.
If $V$ is large, the system takes longer to reach $N_{i}=0$ even if
the rate of reaction and the initial concentrations are the same.
Thus, it is less likely to exhibit switching states
for the same interval of the inflow, $1 / DV$.

In general, some reactions are much faster than the inflow.
If the number of molecules which enter within the time scale of
the reactions is of the order of $1$ for a certain chemical,
the reactions may consume all the molecules of the chemical,
and the molecular discreteness of the chemical becomes significant.
In other words, for the effect of the discreteness to appear
in a system with several processes,
it is important that the number of events of a process within
the time scale of another process is of the order of $1$.

Once the system is in the switching state,
it is fairly stable and difficult to escape,
especially if $DV$ is small.
To escape the 1-3 rich state and regain continuous vibration,
at least one $X_{2}$ molecule and one $X_{4}$ molecule should flow in
and coexist.
It is required that after an $X_{2}$ molecule flows in,
an $X_{4}$ molecule should flow in before $N_{2}$ returns to $0$,
or vice versa.

\begin{figure}
\begin{center}
\includegraphics[width=69mm]{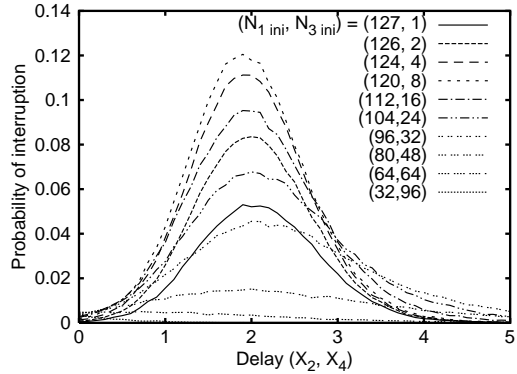}
\end{center}
\caption{Probability for interruption of the 1-3 rich state
as a function of the delay $\tau$,
sampled over $10^{5}$ times for each condition.
We assume the system where $N_{2}=N_{4}=0$.
We inject an $X_{2}$ molecule at $t=0$,
then an $X_{4}$ molecule at $t=\tau$ (no further flow).
If $\forall i : N_{i}>0$ at $t=8$, we ascertain that the system
has escaped from the 1-3 rich state.
$V=32$, $N_{1 ini}+N_{3 ini}=4V (=128)$.
An initial large imbalance, such as $N_{1 ini}:N_{3 ini} = 120:8$ ($15:1$), makes
it easier to escape the 1-3 rich state.
The system is unlikely to escape from the state
when $N_{1 ini}\approx N_{3 ini}$ or $N_{1 ini} < N_{3 ini}$.
The probability is maximum at $\tau\approx 2$,
which approximately corresponds to a half of the period of the vibration
around the fixed point.}
\label{fig:A-sw3b-32}
\end{figure}

We put one $X_{2}$ molecule into the system in the 1-3 rich state
at $t=0$, and one $X_{4}$ molecule in at $t=\tau$.
We assume that $DV$ is so small that no more molecules flow in or out.
Then, we judge whether the system escapes from the 1-3 rich state.
In due course, the system returns to the 1-3 rich
(or sometimes 2-4 rich) state because there is no further flow;
thus, we should judge at the right moment.
Here, if $N_{i} > 0$ for all $i$ at $t=8$ (i.e., waiting about 2.5 times
longer than the period of the oscillation around the fixed point),
we consider that the 1-3 rich state has been interrupted.
We measure the probability of interruption for
various initial conditions and the delay $\tau$ as shown in Fig. \ref{fig:A-sw3b-32}.

The system requires adequate timing of inflow to escape
the switching states, which may amplify the imbalance
between the stability of each state.
For example, to escape the 1-3 rich, $N_{1} > N_{3}$ state,
it is required that an $X_{2}$ molecule flows in, and
then an $X_{4}$ molecule flows in with a certain delay $\tau$,
as shown in Fig. \ref{fig:A-sw3b-32}.
Thus, the frequency of escape from the 1-3 rich state is approximately
proportional to the product of the inflow frequencies of $X_{2}$ and $X_{4}$.
If each $D_{i}$ or $s_{i}$ is species-dependent,
the stability of the 1-3 and 2-4 rich states may strongly depend on
$D_{i}s_{i}V$, the inflow frequency of $X_{i}$.

Furthermore, if at the outset $N_{1} \approx N_{3}$, it is difficult
for the system to escape from the 1-3 rich state.
In some cases where the parameters are species-dependent,
the flows or the switching may lead to $N_{1} \approx N_{3}$,
which stabilizes the 1-3 rich state.

These conditions are important to stabilize particular states and
affect the macroscopic behavior of the system (see Section \ref{sect:tr2}).

\subsection{Stability and Shape of the Network: $k \ne 4$ case}

\begin{figure}
\begin{center}
\includegraphics[width=69mm]{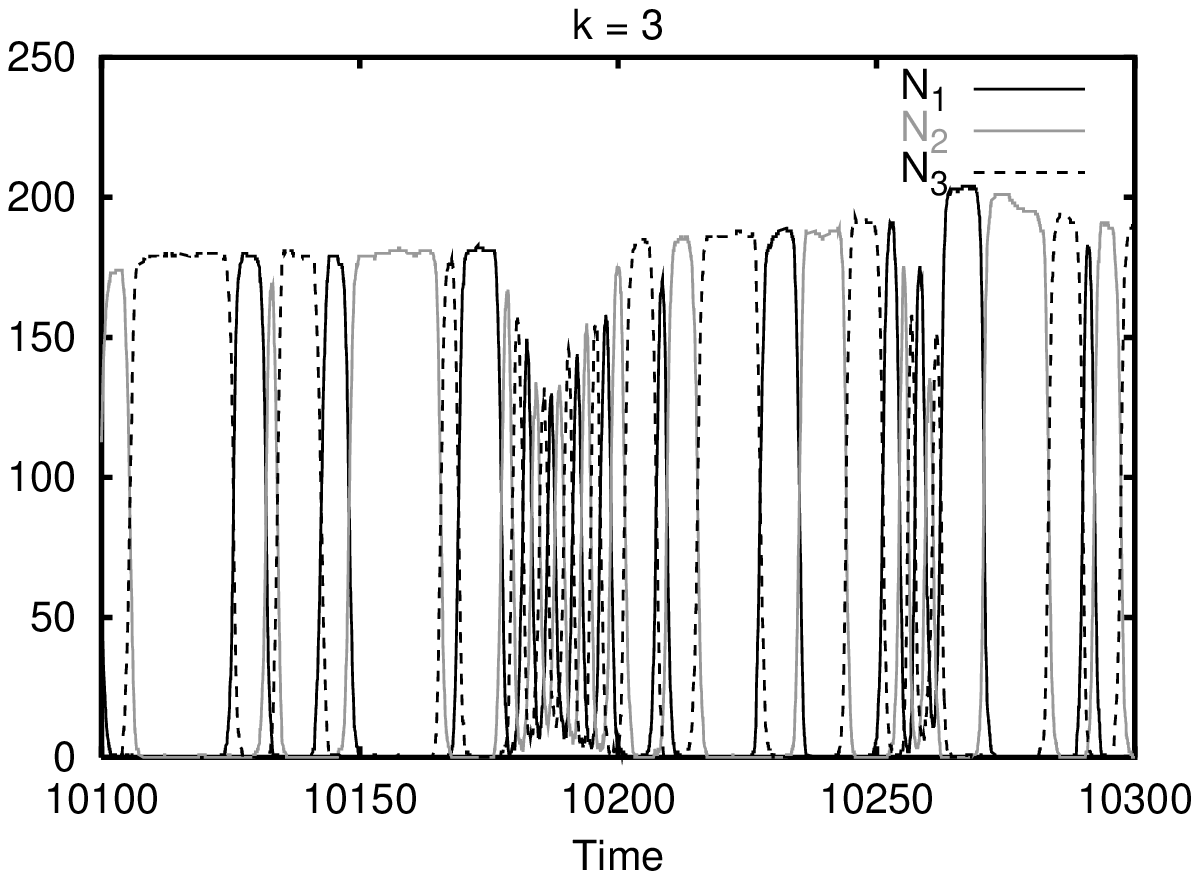}
\includegraphics[width=69mm]{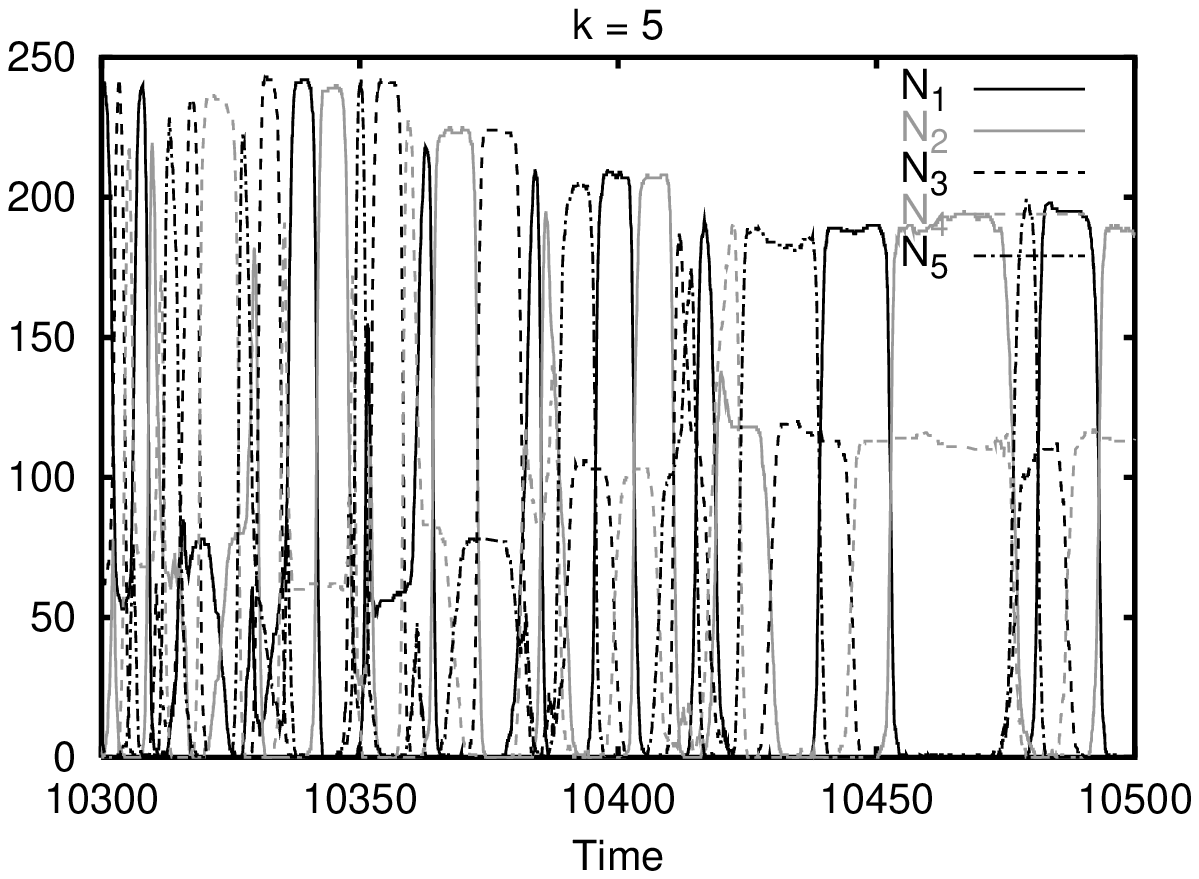}
\includegraphics[width=69mm]{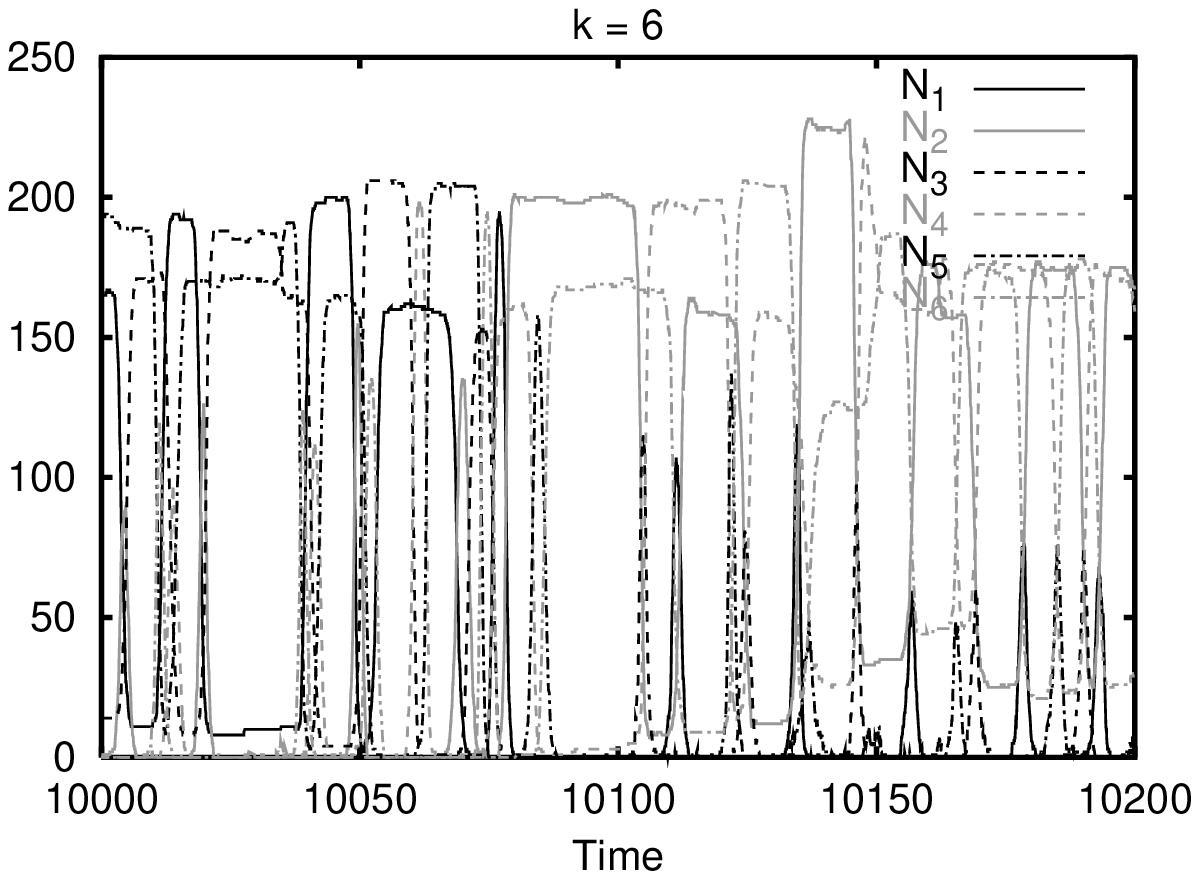}
\end{center}
\caption{Time series of $N_{i}$ for $k=3$, $5$, $6$,
with $V=64$ and $D=1/256$.
For the case where $k=6$, the transition occurs
from the 1-3-5 rich state to the 2-4-6 rich state
at $t \approx 10080$.
For the cases where $k=3$ and $k=5$, there are no stable states
such as the 1-3 rich state when $k=4$ or 1-3-5 rich state when $k=6$.}
\label{fig:A-k356-16}
\end{figure}

To close the section \ref{sect:tr1},
we briefly discuss the cases where $k \ne 4$.
Figure \ref{fig:A-k356-16} shows the time series of each $N_{i}$ for
$k=3$, $5$, and $6$.
When $k=6$, 1-3-5 rich and 2-4-6 rich states appear.
However, these states are less stable than the 1-3 and 2-4 rich states for $k=4$.
These states collapse when any of the rich chemicals vanish;
thus, they are unstable for large $k$.
When $k$ is odd ($3$, $5$, $\cdots$), there are no stable states where
particular chemicals are extinct.
However, additional reactions, or a variety of $r_{i}$ or $s_{i}$,
may stabilize or destabilize the states
such that it is not always true that loops with odd-$k$ are unstable.

The discreteness-induced transitions are
not limited in the autocatalytic loop.
We apply the abovementioned discussions to certain segments of
a complicated reaction network with a slight modification.


\section{Alteration of Concentrations by the DIT}\label{sect:tr2}

In the preceding section, we show that the discreteness of the molecules
can induce transitions to novel ``switching'' states
in autocatalytic systems.

For the case where $k=4$ with uniform parameters,
the 1-3 rich state and the 2-4 rich state are equivalent.
In due course, the system alternates between the 1-3 rich and 2-4 rich states.
The long-term averaged concentrations are still the same as
the continuum limit value, $\bar{x_{i}} = 1$.

It will be important if macroscopic properties, such as the average
concentrations, can be altered.
We show that the discreteness-induced transitions may alter
the long-term averaged concentrations.

\subsection{Model}

Once again, here, we adopt the autocatalytic reaction loop
\[
X_{i} + X_{i+1} \rightarrow 2X_{i+1}
\]
for the $k=4$ species.
Now we consider the case where the parameters $D_{i}$, $s_{i}$, or $r_{i}$ are
species-dependent.
In the continuum limit, the concentration $x_{i}$ is governed by
the rate equation
\begin{equation}
\frac{dx_{i}}{dt} = r_{i-1}x_{i-1}x_{i} - r_{i}x_{i}x_{i+1} + D_{i}(s_{i} - x_{i}) .
\end{equation}
The rate equation does not contain the volume $V$;
hence, the average concentrations should be independent of $V$.

As discussed in the preceding section, for the transitions
to the switching states to occur, it is necessary
that the interval of the inflow is longer than
the time scale of the reactions.
In this model, the inflow interval of $X_{i}$ is $\sim 1/D_{i}s_{i}V$, and
the time scale of the reaction $X_{i} + X_{i+1} \rightarrow 2X_{i+1}$ in order to
use $X_{i}$ up is $\sim 1/r_{i}x_{i+1}$.
If all the chemicals are equivalent,
the discreteness of all the chemicals equally take effect,
and the 1-3 and 2-4 rich states coordinately appear at $DV \approx r$.

Now, since the parameters are species-dependent,
the effect of discreteness may be different for each species.
For example, assuming that $D_{1}s_{1} < D_{2}s_{2}$,
the inflow interval of $X_{1}$ is longer than that of $X_{2}$.
Thus, the discreteness of the inflow of $X_{1}$ may be significant for larger $V$.

To demonstrate a possible effect of the discreteness on the macroscopic properties
of the system,
we measure each average concentration $\bar{x_{i}}$,
sampled over a long enough time to allow transitions between
the 1-3 and 2-4 rich states, by Gillespie's direct method.
Note that every $\bar{x_{i}}$ does not depend on $V$ in the continuum limit.
Generally, in discrete simulations, the effect of the discreteness varies
with $V$ and alters every $\bar{x_{i}}$.
When $V$ is very large, the discreteness does not matter
and $\bar{x_{i}}$ is almost equal to the continuum limit value.
In contrast, when $V$ is small,
the discreteness causes $\bar{x_{i}}$ to be very different
from the continuum limit.

We first investigate the case where each $s_{i}$ is species-dependent
(i.e., each inflow rate is species-dependent),
and each $D_{i}$ and $r_{i}$ are uniform ($D_{i}=D$, $r_{i}=1$).
Later, we briefly discuss the case where $r_{i}$ is inhomogeneous.

As mentioned in Section \ref{subsect:tr1-cond},
for the effect of the discreteness to appear,
it is important that the interval of events of a process is
of the order of or longer than the time scale of another process.
As regards the inflow,
there are two indices to determine how the discreteness appears:
\begin{itemize}
\item The inflow interval, $\frac{1}{Ds_{i}V}$,
\item The number of molecules at equilibrium, $s_{i}V$.
\end{itemize}

If the inflow interval, $\frac{1}{Ds_{i}V}$, is longer than the time scale
of the reactions,
the reactions may exhaust the chemical before the chemical enters,
and the inflow discreteness becomes significant.

Furthermore, if $s_{i}V$ is smaller than $1$,
$N_{i}$ can reach $0$ because of the outflow.
In such cases, the relation between the inflow interval and
the outflow time scale is also important.
The approach time from $N_{i}=n \gg 1$ to $N_{i}=0$ is
of the order of $\frac{1}{D} \log n$.
If the inflow interval of the chemical that causes the switching to raise $N_{i}$ is
long enough to allow all $X_{i}$ molecules to flow out,
the inflow discreteness may alter the stability of the states drastically.

From this point of view,
we classify the mechanism in cases I, I$'$, and II as follows.

\subsection{Case I: inflow discreteness and reaction rate}

\begin{figure}
\begin{center}
\includegraphics[width=69mm]{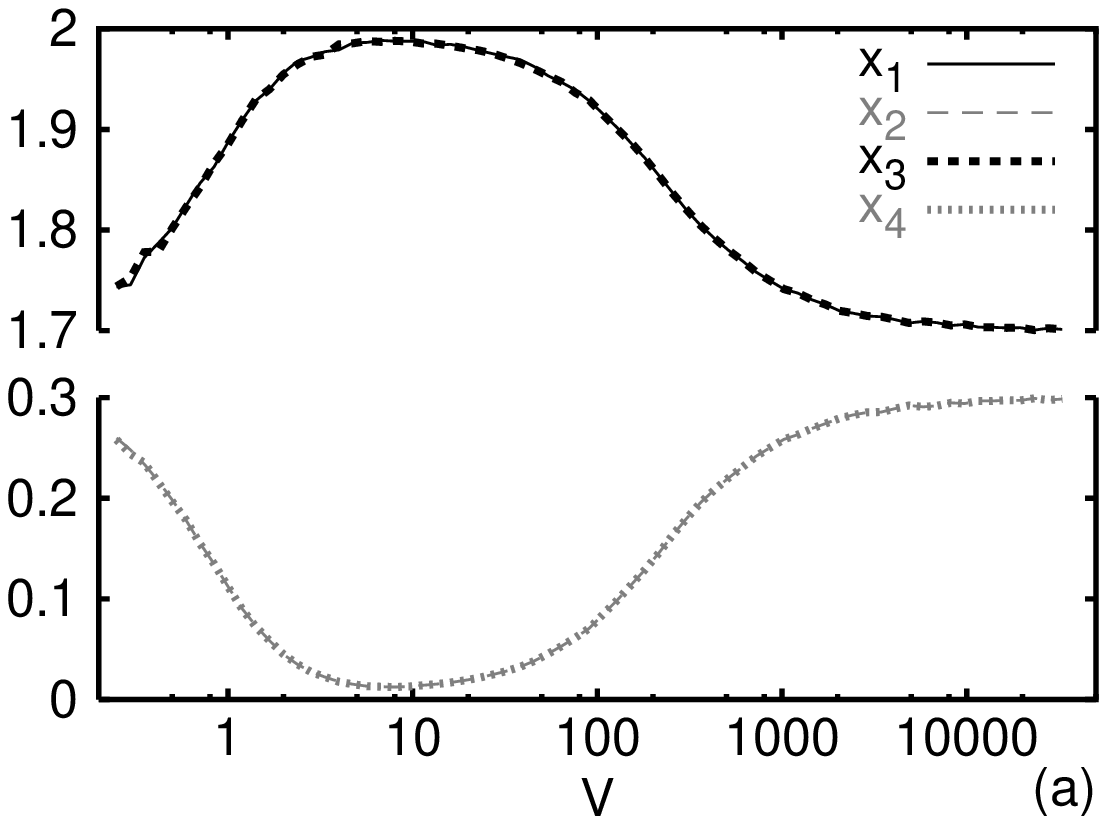}
\includegraphics[width=69mm]{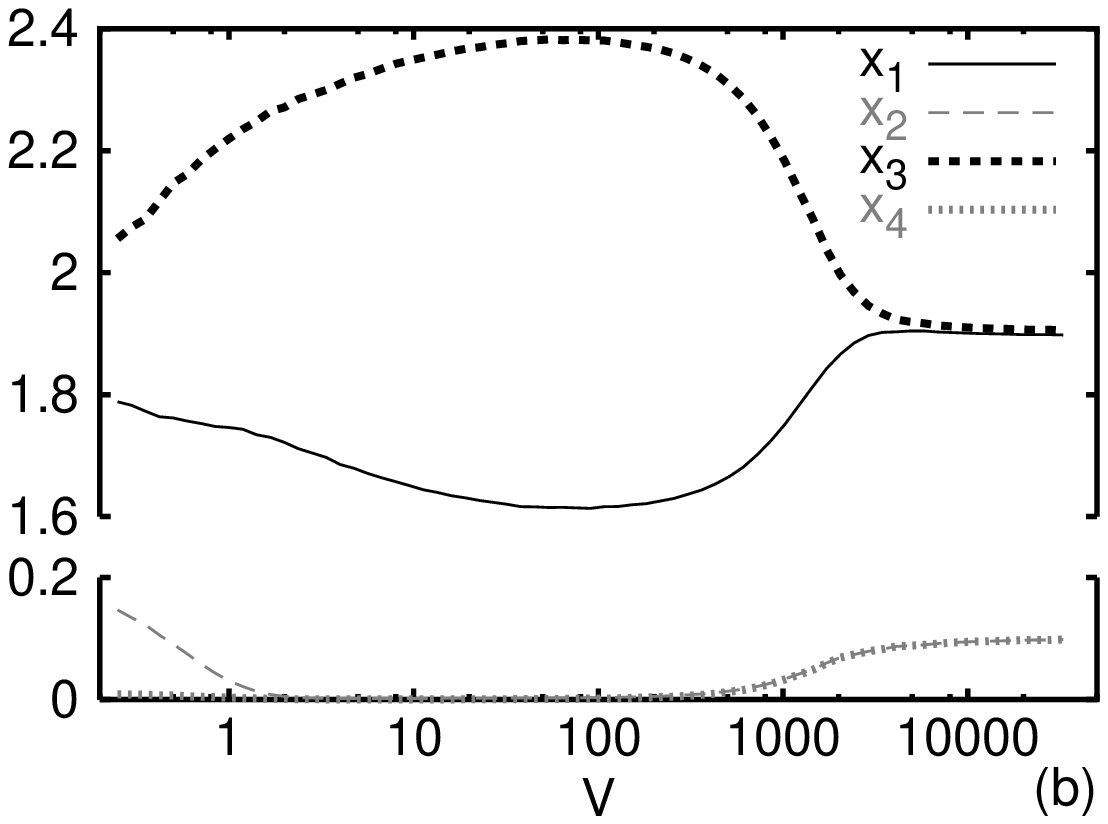}
\end{center}
\caption{The average concentration $\bar{x_{i}}$ as a function of $V$,
sampled over $10^{6}$ ($V>1024$), $10^{7}$ ($32<V\le 1024$),
and $10^8$ ($V\le 32$) time units (same for Fig. \ref{fig:tr2-average2}).
$r=1$ and $D=1/128$.
(a) Case I: $s_{1}=s_{3}=1.7$, $s_{2}=s_{4}=0.3$.
When $V$ is small, the 1-3 rich state appears, and
the difference between $\bar{x_{1}}$ and $\bar{x_{2}}$ increases.
(b) Case I$'$: $s_{1}=s_{3}=1.9$, $s_{2}=0.19$, $s_{4}=0.01$.
When $V$ is small, $\bar{x_{2}}$ and $\bar{x_{4}}$ decrease, identical to Case I.
In this case, the imbalance between $\bar{x_{1}}$ and $\bar{x_{3}}$ appears
at the same time.
(Figures \ref{fig:tr2-average1}, \ref{fig:tr2-average2}, \ref{fig:tr2-average3},
and \ref{fig:tr2-dist-x2} are reproduced from ref. \cite{Togashi2}
by permission of the publisher.)
}
\label{fig:tr2-average1}
\end{figure}

We start with the simplest case, $s_{1} = s_{3} > s_{2} = s_{4}$.
In this case, the rate equation has a stable fixed point
with $\forall i: x_{i} = s_{i}$.
When $V$ is large, each $x_{i}$ fluctuates around the fixed point,
and each average concentration $\bar{x_{i}}$ is in accordance
with the fixed point value.
When $V$ is small, $\bar{x_{i}}$ depends on $V$.
Fig. \ref{fig:tr2-average1} shows each $\bar{x_{i}}$ as a function of $V$.
The difference between $\bar{x_{1}}$ and $\bar{x_{2}}$ increases
for small $V$.

By decreasing $V$, first $N_{2}$ and $N_{4}$ reach $0$ and
the 1-3 rich state appears.
To reach $N_{2}=0$ or $N_{4}=0$, the inflow interval of $X_{2}$ or $X_{4}$ should be
longer than the time scale of the reactions.
We set $x_{i} = O(1)$;
thus, the condition to achieve the 1-3 rich state is approximately
$\frac{1}{Ds_{2}V}$, $\frac{1}{Ds_{4}V} > \frac{1}{r}$;
that for the 2-4 rich state is $\frac{1}{Ds_{1}V}$, $\frac{1}{Ds_{3}V} > \frac{1}{r}$.

If $V$ satisfies
$\frac{1}{Ds_{1}V}$, $\frac{1}{Ds_{3}V} < \frac{1}{r} < \frac{1}{Ds_{2}V}$, $\frac{1}{Ds_{4}V}$,
the 1-3 rich state appears but the 2-4 rich state does not.
Thus, $\bar{x_{1}}$ and $\bar{x_{3}}$ increase.
We actually observed this at $V \approx \frac{r}{Ds_{2}}$,
as shown in Fig. \ref{fig:tr2-average1}.

For smaller $V$ that fulfills both the 1-3 and 2-4 rich states,
the imbalance between the 1-3 and 2-4 rich states does not disappear.
Once the system is in the 1-3 rich state,
adequate timing for the $X_{2}$ and $X_{4}$ inflow is required
to escape the state.
Thus, the frequency of escape from the 1-3 rich state
is approximately proportional to $D^{2}s_{2}s_{4}V^{2}$,
the product of the inflow frequencies of $X_{2}$ and $X_{4}$.
The average residence time in the 1-3 rich state
as well as the 2-4 rich state is
the reciprocal of the escape frequency.
The ratio of the average residence time in the 1-3 rich state
to that in the 2-4 rich state is $\approx \frac{s_{1}s_{3}}{s_{2}s_{4}}$.
In addition, after escaping the switching states,
the system tends to reach the 1-3 rich state rather than the 2-4 rich state
because of the biased inflow.
Thus, the ratio of the total residence time in the 1-3 rich state
to that of the 2-4 rich state is larger than $\frac{s_{1}s_{3}}{s_{2}s_{4}}$;
hence, $\bar{x_{1}}$, $\bar{x_{3}} \gg \bar{x_{2}}$, $\bar{x_{4}}$, even for
a small difference in $s_{i}$.

\subsection{Case I$'$: imbalance of inflow discreteness}

In Case I, we consider the imbalance between the $X_{1}$, $X_{3}$ pair
and the $X_{2}$, $X_{4}$ pair.
If another imbalance exists between the $X_{2}$ and $X_{4}$ inflows,
the switching induced by these chemicals in the 1-3 rich state may be unbalanced.
We consider the case where $s_{1} = s_{3} > s_{2} > s_{4}$.

In this case, the 1-3 rich state is more stable than the 2-4 rich state,
which is identical to Case I.
In the 1-3 rich state, the system can be switched
from $N_{1} > N_{3}$ to $N_{1} < N_{3}$ by an $X_{2}$ molecule;
and from $N_{1} < N_{3}$ to $N_{1} > N_{3}$ by an $X_{4}$ molecule.
Now, the inflow rate of $X_{2}$ is larger than $X_{4}$; thus,
switching from  $N_{1} > N_{3}$ to $N_{1} < N_{3}$ is more probable
than vice versa, and the system tends to stay in the $N_{1} < N_{3}$ state.
Consequently, $\bar{x_{1}} < \bar{x_{3}}$, as shown in Fig. \ref{fig:tr2-average1}.
This effect requires switching states.
When $V$ is large, $\bar{x_{1}}$ and $\bar{x_{3}}$ are almost the same.

\subsection{Case II: inflow and outflow}

Here, we consider the case where $s_{1} = s_{2} > s_{3} = s_{4}$.
In this case also, the rate equation has a stable fixed point\footnote{Generally, if $D \ll rs_{i}$,
$x_{1}$, $x_{3} \approx (s_{1} + s_{3}) / 2$ and
$x_{2}$, $x_{4} \approx (s_{2} + s_{4}) / 2$.}.

When $V$ is small, both the 1-3 and the 2-4 rich states appear.
Here, we consider $s_{4}V$, the number of $X_{4}$ molecules
when the concentration of $X_{4}$ in the container and
in the reservoir are at equilibrium.
If $s_{4}V < 1$, $N_{4}$ reaches $0$ without undergoing any reaction.
The system takes $\sim \frac{1}{D} \log n$ time units to reach from $N_{4}=n \gg 1$ to $N_{4}=0$.
The reaction $X_{4} + X_{1} \rightarrow 2X_{1}$ also uses $X_{4}$ such that
$N_{4}$ decreases faster if $s_{1}$ is large.

If $N_{4} > 0$, the inflow of $X_{3}$ may switch from $N_{2} > N_{4}$ to
$N_{2} < N_{4}$ and raise $N_{4}$ again.
The inflow interval of $X_{3}$ is $\sim \frac{1}{Ds_{3}V}$.
If the interval is much shorter than the approach time to $N_{4}=0$,
the switching maintains $N_{4} > 0$.
If the interval is longer, $N_{4}$ reaches $0$ before switching,
and the 2-4 rich state is easily destroyed.

In the 1-3 rich state, the system tends to maintain $N_{1} < N_{3}$ because
the $X_{2}$ inflow is frequent.
However, the $X_{1}$ inflow is also large enough to maintain $N_{1} > 0$.
The 1-3 rich state retains its stability.

\begin{figure}
\begin{center}
\includegraphics[width=69mm]{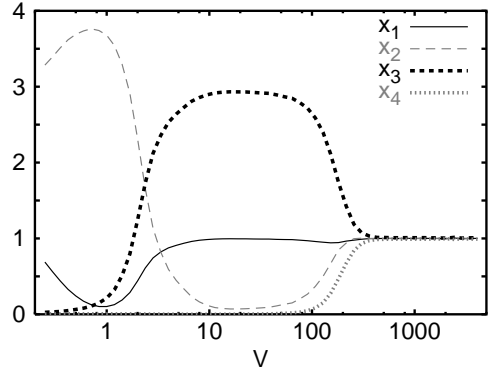}
\end{center}
\caption{The average concentration $\bar{x_{i}}$, for
Case II: $s_{1}=s_{2}=1.99$, $s_{3}=s_{4}=0.01$, $r=1$, and $D=1/128$.
For small $V$, the 1-3 rich state is preferred, and $\bar{x_{3}}$ increases.}
\label{fig:tr2-average2}
\end{figure}

In conclusion, the 1-3 rich state is more stable than the 2-4 rich state.
In the 1-3 rich state, $N_{1} < N_{3}$ is preferred, and
$\bar{x_{3}}$ increases.
At this stage, it is possible that
$\bar{x_{2}} \ll \bar{x_{3}}$ despite the fact that $s_{2} \gg s_{3}$,
as shown in Fig. \ref{fig:tr2-average2}.

\subsection{Amplification by Discreteness}

\begin{figure}
\begin{center}
\includegraphics[width=69mm]{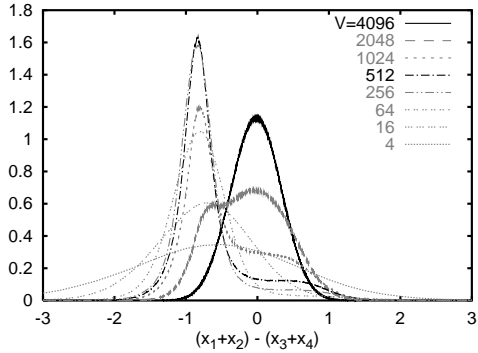}
\end{center}
\caption{Probability distribution of $y \equiv (x_{1}+x_{2})-(x_{3}+x_{4})$, for
Case I$'$: $s_{1}=s_{3}=1.9$, $s_{2}=0.19$, $s_{4}=0.01$, $r=1$, and $D=1/128$.
There is a peak around $y=0$ for large $V$.
The difference in $s_{i}$ has a slight effect on $\bar{x_{i}}$.
For the case where $V\le 1024$, there appears another peak at $y\approx -0.8$,
which corresponds to the 1-3 rich, $N_{1}<N_{3}$ state.}
\label{fig:tr2-dist-ab-cd-caseI'}
\end{figure}

In summary, the difference in the ``extent of discreteness'' between
chemical species induces novel transitions.
The ``extent of discreteness'' depends on $V$; thus, we observe
transitions by changing $V$.
The transition reported in Section \ref{sect:tr1} is
regarded as a {\sl second order} transition involving symmetry breaking
(see Figs. \ref{fig:A-ac-bd} and \ref{fig:A-ab-cd}),
while the transition in this section corresponds to
the {\sl first order} transition without symmetry breaking
(see Fig. \ref{fig:tr2-dist-ab-cd-caseI'})
in terms of thermodynamics.

\begin{figure}
\begin{center}
\includegraphics[width=69mm]{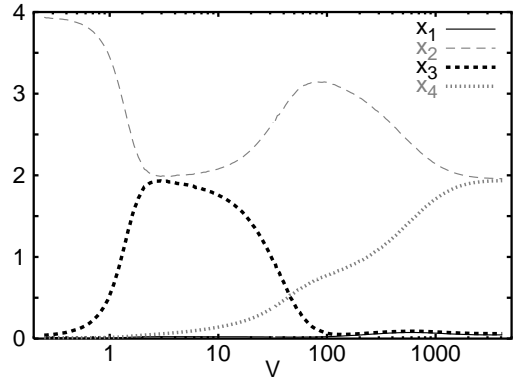}
\end{center}
\caption{The average concentration $\bar{x_{i}}$ for
$s_{1}=0.09$, $s_{2}=3.89$, $s_{3}=s_{4}=0.01$, $r=1$, and $D=1/64$,
sampled over $5 \times 10^{6}$ ($V > 32$) or
$5 \times 10^{8}$ ($V \le 32$) time units.
By decreasing $V$, first, the 2-4 rich state appears, as seen in Case I.
Then, the 2-4 rich state becomes unstable
and gives way to the 1-3 rich state, as seen in Case II.
When $V$ is extremely small ($V<0.5$),
the flow of molecules governs the system, and $\bar{x_{2}}$ increases again.}
\label{fig:tr2-average3}
\end{figure}

We classified the mechanism in cases I, I$'$, and II.
These mechanisms can be combined.
For example, we demonstrate the case where $s_{1}=0.09$, $s_{2}=3.89$, and $s_{3}=s_{4}=0.01$.
In this case, each $\bar{x_{i}}$ shows a three-step change with $V$, as shown
in Fig. \ref{fig:tr2-average3}.

\begin{figure}
\begin{center}
\includegraphics[width=69mm]{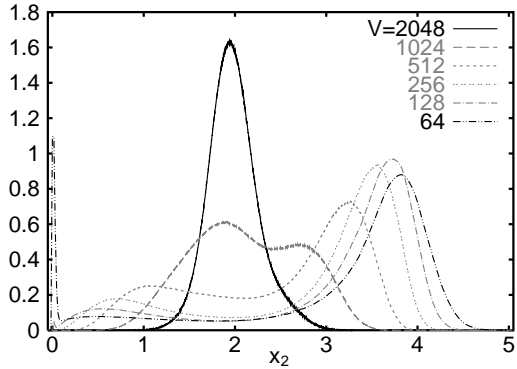}
\end{center}
\caption{Distribution of $x_{2}$ for
$s_{1}=0.09$, $s_{2}=3.89$, $s_{3}=s_{4}=0.01$, $r=1$, and $D=1/64$,
sampled over $5 \times 10^{6}$.
For large $V$, a single peak around $x_{2}=2$ appears,
which corresponds to the fixed point in the continuum limit.
At $V \approx 10^{3}$, double peaks appear around
$x_{2}=1$ and $x_{2}=3$, which correspond to the 2-4 rich state.
By decreasing $V$, the two peaks spread apart.
At $V \approx 10^{2}$, the skirt of the low-density (left) peak
touches $x_{2}=0$, implying that $N_{2}$ (and $N_{4}$) is likely to
reach $0$, and thus, the 2-4 rich state loses stability.
A peak at $x_{2}=0$ steeply grows by decreasing $V$ further.}
\label{fig:tr2-dist-x2}
\end{figure}

When $V$ is large, $\bar{x_{1}}$, $\bar{x_{3}} \ll \bar {x_{2}}$, $\bar{x_{4}}$,
since $s_{1}+s_{3} \ll s_{2}+s_{4}$.
At $V \approx 10^{3}$, the discreteness of the $X_{3}$ inflow becomes significant,
and the 2-4 rich state appears.
In the 2-4 rich state, the system tends to remain at $N_{2} > N_{4}$ because of the inflow imbalance
between $X_{1}$ and $X_{3}$, as observed in Case I$'$.
Figure \ref{fig:tr2-dist-x2} shows the distribution of $x_{2}$.
The major peak corresponds to the 2-4 rich, $x_{2} > x_{4}$ state
for the cases when $128 \le V \le 512$.

\begin{figure}
\begin{center}
\includegraphics[width=69mm]{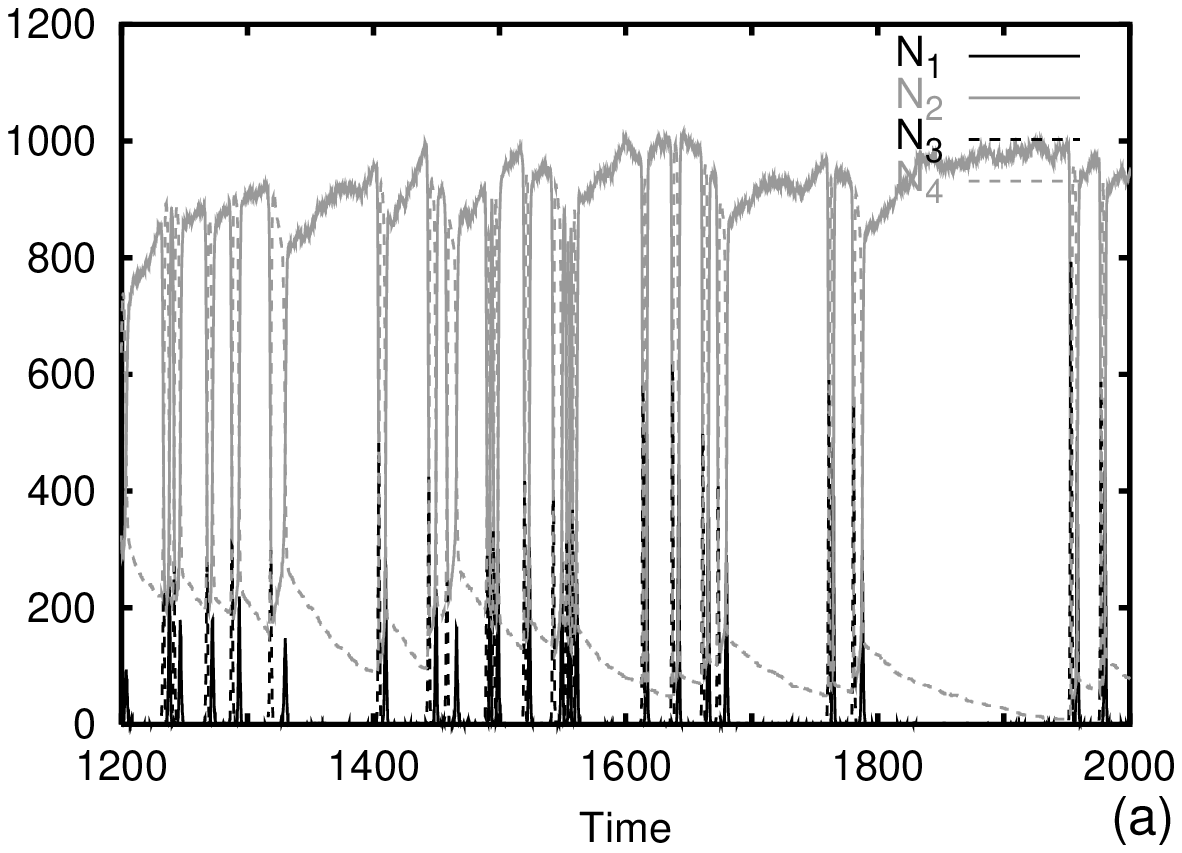}
\includegraphics[width=69mm]{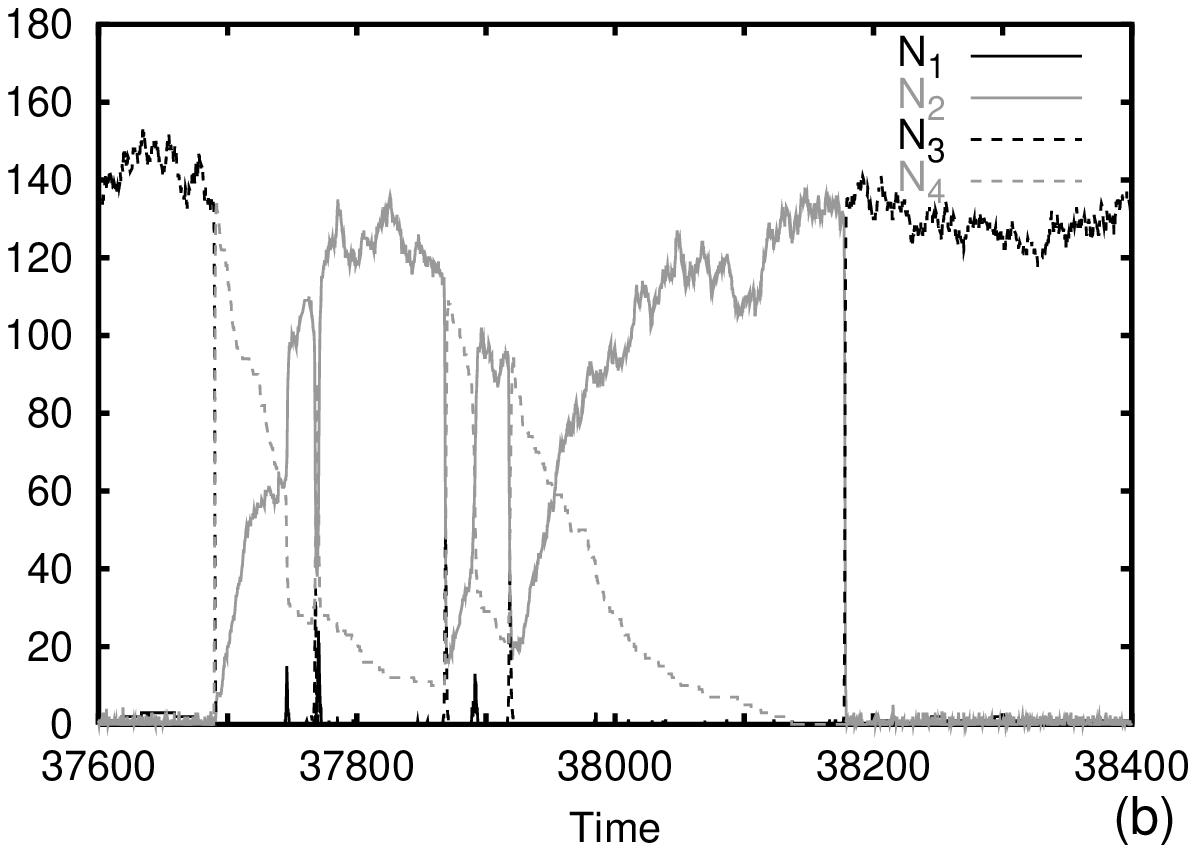}
\end{center}
\caption{Time series of $N_{i}$ for
$s_{1}=0.09$, $s_{2}=3.89$, $s_{3}=s_{4}=0.01$, $r=1$, and $D=1/64$.
(a) $V=256$. The 2-4 rich state is dominant (Case I$'$).
Inflow of $X_{3}$ molecules induces
switching from $N_{2} > N_{4}$ to $N_{2} < N_{4}$,
which prevents $N_{4}$ from decreasing to $0$.
(b) $V=32$. Now, the $X_{3}$ inflow is rare,
which allows $N_{4}$ to reach $0$ before the switching.
Thus, the 2-4 rich state is unstable (Case II).}
\label{fig:tr2-ts-amp}
\end{figure}

On the other hand, in the 2-4 rich state, the outflow of $X_{4}$ depresses
$N_{4}$ toward $s_{4}V$, as observed in Fig. \ref{fig:tr2-ts-amp}.
By decreasing $V$, the imbalance between $N_{2}$ and $N_{4}$ increases
because the rate of switching, which again raises $N_{4}$,
decreases in proportion to $V$.
Finally, at $V \approx 10^{2}$, the 2-4 rich state loses stability,
as seen in Case II.
Now, the 1-3 rich state is preferred despite the fact that $s_{1}+s_{3} \ll s_{2}+s_{4}$.
$\bar{x_{3}}$ increases to $\approx 2$, which is more than $30$ times as large as
that in the continuum limit.

For extremely small $V$, the 1-3 rich state is also unstable because
$N_{1}$ and $N_{3}$ easily reach $0$.
In such a situation, typically only one chemical species is in the container.
The system is dominated by diffusion, and $\bar{x_{2}}$ increases again due to
the large $s_{2}$.

Note that the chemical that becomes extinct depends not only on the flows
but also on the reactions.
In some cases, we observe smaller $\bar{x_{3}}$ for larger $s_{3}$.

\subsection{Asymmetric Reaction Case}

It is possible that each $r_{i}$ varies with the species.
In such cases, we can discuss the effect of discreteness in a similar way.
However, the change of $\bar{x_{i}}$ with $V$ is different from the
case with asymmetric flows.

\begin{figure}
\begin{center}
\includegraphics[width=69mm]{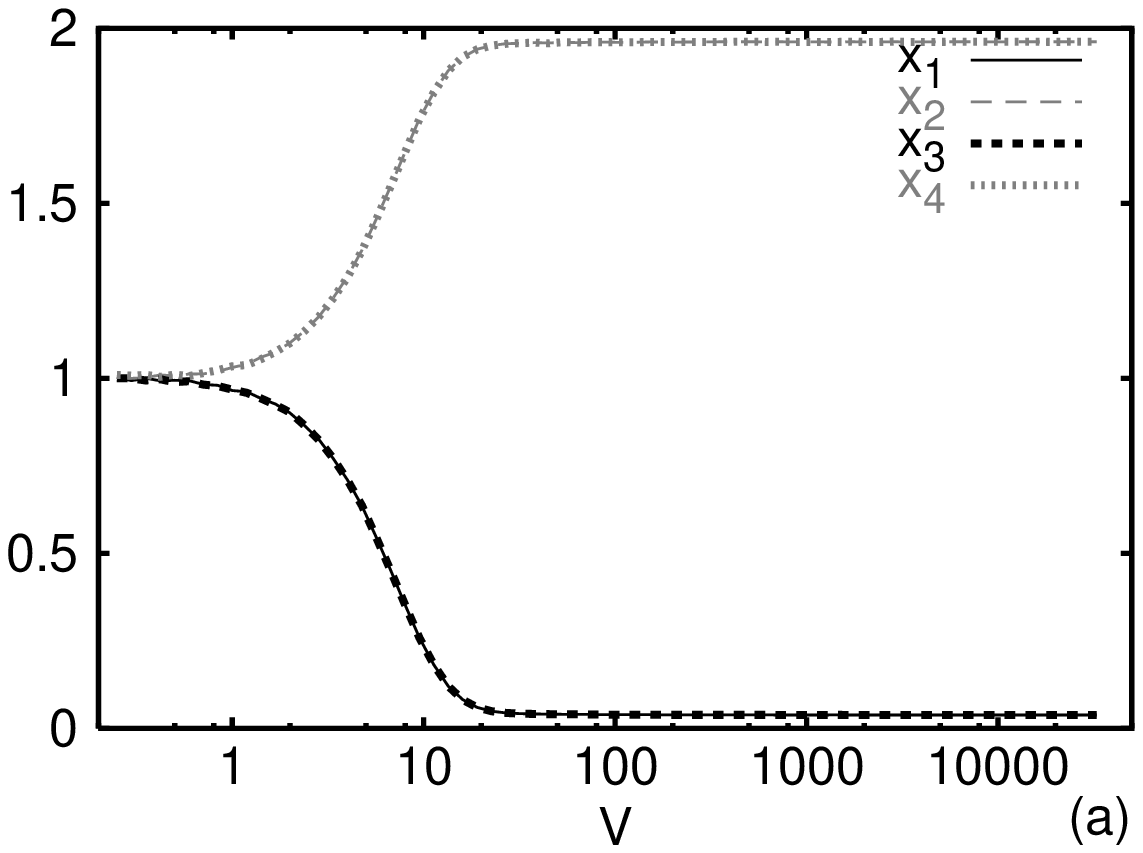}
\includegraphics[width=69mm]{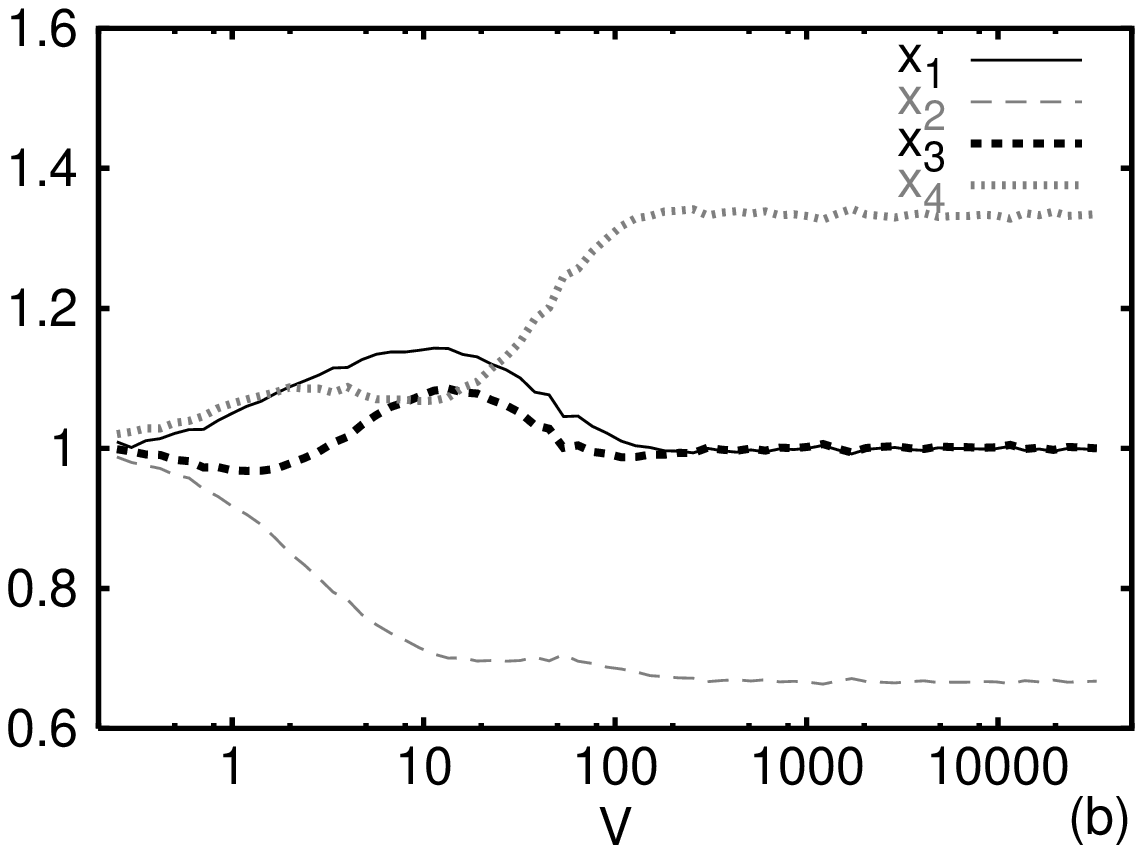}
\end{center}
\caption{The average concentration $\bar{x_{i}}$ for
$\forall i : s_{i}=1$ and $D=1/128$ with
inequivalent reaction constants.
For small $V$, the flows of molecules dominate the system.
Thus, $\bar{x_{i}} \approx 1$, which simply reflects $s_{i}=1$;
this does not depend on how the continuum limit is imbalanced by the reactions.
(a) $r_{1}=r_{3}=1$ and $r_{2}=r_{4}=0.9$.
(b) $r_{1}=r_{2}=2$ and $r_{3}=r_{4}=1$.}
\label{fig:tr2-average-r}
\end{figure}

For example, we assume that $r_{1} = r_{3} > r_{2} = r_{4}$ and
$\forall i: s_{i}=1$.
In the continuum limit or in the case of large $V$,
$\bar{x_{2}} = \bar{x_{4}} > \bar{x_{1}} = \bar{x_{3}}$,
as shown in Fig. \ref{fig:tr2-average-r}.
In contrast, when $V$ is small, $\bar{x_{i}} \approx 1$.
If $V$ is very small, such that the total number of molecules is mostly $0$ or $1$,
reactions rarely take place.
The flow of chemicals dominate the system;
thus, $\bar{x_{i}} \approx s_{i}$.

If both the reactions and the flows are species-dependent,
we simply expect the behavior to be a combination of the abovementioned cases.
Even this simple system can exhibit a multi-step change in concentrations
along with a change in $V$.
It is not limited to the simple reaction loop.
In fact, we observe this kind of change in concentrations
with a change in the system size in randomly connected reaction networks.
For a large reaction network with multiple time scales of reactions
and flows, the discreteness effect may exhibit behavior that is more complicated.
Our discussion is largely applicable to such cases if we can
define the time scales appropriately.

As seen in this paper, the discreteness of molecules can alter the average
concentrations.
When the rates of inflow and/or the reaction are species-dependent,
transitions between the discreteness-induced states are imbalanced.
This may alter the average concentrations drastically
from those of the continuum limit case.


\section{Discussion}

We demonstrated that molecular discreteness may induce transitions to
novel states in autocatalytic systems,
and that may result in an alteration of the macroscopic
properties such as the average chemical concentrations.

In biochemical pathways, it is not anomalous that the number of
molecules of a chemical is of the order of $10^2$ or less
in a cell.
There are thousands of protein species,
and the total number of protein molecules in a cell is not very large.
For example, in signal transduction pathways, some chemicals
work at less than $100$ molecules per cell.
There exist only one or a few copies of genetic molecules such as DNA;
furthermore, mRNAs and tRNAs are not present in large numbers.
Thus, regulation mechanisms involving genes are quite stochastic.
Molecular discreteness naturally concerns such rare chemicals.

One of the authors, Kaneko, and Yomo recently provided the
``Minority Control conjecture,''
which propounds that chemical species with a small number of molecules
governs the behavior of a replicating system,
which is related to the origin of heredity \cite{KanekoYomoMinority,KanekoMinNet2002,KanekoMinNet2003}.
Matsuura et al. experimentally demonstrated that
a small number of genetic molecules is essential for evolution \cite{MinorityExperiment}.
Molecular discreteness should be significant for such chemicals,
and may be relevant to characters of genetic molecules.

Until now, we have modeled reactions in a well-stirred medium,
where only the number of molecules is taken into account while determining the behavior.
However, if the system is not mixed well, we should take into account
the diffusion in space.
Both the total number of molecules and the spatial distributions
of the molecules may be significant.
From a biological point of view, the diffusion in space is also important
because the diffusion
in cells is not always fast as compared with the time scales of the reactions.
If the reactions are faster than the mixing, we should consider the system
as a reaction-diffusion one, with discrete molecules diffusing in space.
The relation between these time scales will be important,
as indicated by Mikhailov and Hess \cite{Mikhailov2002,Mikhailov1A,Mikhailov1B}.
As regards these time scales, we recently found that the spatial discreteness
of molecules within the so-called Kuramoto length \cite{Kampen,Kuramoto1,Kuramoto2},
over which a molecule diffuses in its lifetime
(lapses before it undergoes reaction),
may yield novel steady states that are not observed
in the reaction-diffusion equations \cite{Togashi3}.
There is still room for exploration in this field,
e.g., pattern formation.

Our result does not depend on the details of the reaction
and may be applicable to systems beyond reactions,
such as ecosystems or economic systems.
The inflow of chemicals in a reaction system can be seen
as a model of intrusion or evolution in an ecosystem;
both systems with discrete agents (molecules or individuals), which
may become extinct.
In this regard, our result is relevant to studies of ecosystems,
e.g., extinction dynamics with a replicator model by Tokita and Yasutomi
\cite{Tokita1999,Tokita2003}.
The discreteness of agents or operations might also be relevant to
some economic models, e.g., artificial markets.

Most mathematical methods that are applied to reaction systems
cannot account for the discreteness.
Although the utility of simulations have become convenient with the progress
of computer technology,
it might be useful if we could construct a theoretical formulation
applicable to discrete reaction systems.
On the other hand, in recent years, major advances have been made
in the detection of a small number of molecules
and fabrication of small reactors,
which raises our hopes to demonstrate the effect of discreteness
experimentally.

We believe that molecular discreteness is
of hidden but real importance with respect to biological mechanisms, such as
pattern formation, regulation of biochemical pathways, or evolution,
to be pursued in the future.

\begin{acknowledgments}
This research is supported by grant-in-aid for scientific research from
the Ministry of Education, Culture, Sports, Science and Technology of Japan
(11CE2006, 15-11161).
Y.T. is supported by a research fellowship from
the Japan Society for the Promotion of Science.
\end{acknowledgments}

\section*{Appendix: Methods for Simulating Discrete Reaction Systems}

Since the 1970s, several methods have been suggested
for simulating discrete reaction systems.
We briefly review some of these methods:
StochSim method, Gillespie's methods, and their improved versions.

These methods are based on chemical master equations.
In chemical master equations, we define the state of the system
as the number of molecules in each chemical;
the reaction process as a series of transitions between the states\footnote{Gillespie demonstrated that the chemical master equations are exact
for gas-phase, well-stirred systems in thermal-equilibrium \cite{Gillespie1992}.}.
We consider each event, i.e., a reaction event between molecules,
inflow or outflow of a molecule, as a transition.
They take place stochastically with a certain frequency (probability per time unit)
determined by the current state.

For the simulations used in this study,
we adopted Gillespie's direct method for simplicity\footnote{The next reaction method would be more suitable
if the system contains more chemical species and reactions.}.
We also attempted a direct simulation and the next reaction method,
and confirmed that our result does not depend on the simulation method.

\subsection{Direct Simulation with Fixed Time Step}

First, let us consider a very simple approximation,
which looks similar to the Euler method for differential equations.

In this method, we fix the time step as $\Delta t$.
Assuming that the frequency of the event-$i$ is $a_{i}$,
the average number of the event-$i$ for each step is $a_{i} \Delta t$.
If $\sum a_{i} \Delta t \ll 1$, we approximately assume
that at most one event occurs at each step,
and the probability of the event-$i$ is $a_{i} \Delta t$ (it is
possible that no event occurs in the step).
We select an event with a random number, change the state
according to the event, and recalculate each $a_{i}$.

\subsection{StochSim Method}

From this simple concept, we can derive some variations.
The so-called StochSim method \cite{MortonFirth1998} is one such variation.

The StochSim method adopts random sampling of molecules.
For bimolecular reactions,
we randomly choose two molecules from the system,
and decide whether they react or not with certain probability.
The method requires three random numbers in total for each step.
In case there are some single molecular (first-order) reactions,
we choose the second molecule
from the molecules in the system and some pseudo-molecules (dummies).
If the second molecule is a pseudo-molecule,
then we select the single molecular reaction of the first molecule,
and determine whether it occurs.

In the StochSim method, $\Delta t$ is restricted by the fastest reaction
(with the largest reaction rate per pair of molecules).
If most of the bimolecular pairs do not react with each other,
or the reaction probability varies with the species,
this method may be impractical because no reaction occurs in most steps.

\subsection{Gillespie's Direct Method}

Incidentally, if the system consists of discrete molecules,
it is typical that each frequency $a_{i}$ changes
only when an event actually occurs.
Taking this into account,
Gillespie suggested two exact simulation methods:
the Direct Method \cite{Gillespie2} and the First Reaction Method \cite{Gillespie1}.
In these methods, we do not fix the time step.
Instead, we calculate the time lapse until the next event.

In the direct method, first,
we consider the total frequency of the events, $a = \sum a_{i}$.
If $a_{i}$ does not change until the next event,
the time lapse until the next event, $\tau$, is
exponentially distributed as $P(\tau) = a e^{-a\tau}$ ($0 < \tau$).
We determine the time lapse $\tau$ with an exponentially distributed
random number.

Subsequently, the probability that the next event is $i$ is $a_{i} / a$.
We determine which event occurs with a random number.
We then set the time $t$ forward by $\tau$,
update the state according to the event,
and recalculate each frequency $a_{i}$.
Iterate the above steps until the designated time elapses.

\subsection{Gillespie's First Reaction Method}

The first reaction method is similar to the direct method.
It is based on the fact
that $\tau_{i}$, the time lapse until the next event-$i$,
is exponentially distributed as
$P(\tau_{i}) = a_{i} e^{-a_{i}\tau_{i}}$ ($0 < \tau_{i}$).
Only the event with minimum $\tau_{i}$ actually occurs.
We update the state, recalculate each $a_{i}$,
and generate all $\tau_{i}$ again with the new corresponding $a_{i}$.

In the first reaction method,
we need as many random numbers each step as the types of events.
We calculate all $\tau_{i}$, choose the earliest, and discard the others.
Generally, the processor time to generate random numbers is very large;
hence, a large amount of time is wasted for several types of events.

\subsection{Next Reaction Method}

To solve this performance problem,
Gibson and Bruck proposed a refinement of the first reaction method:
Next Reaction Method \cite{Gibson}.
Although, in general, Monte Carlo simulations require independency of
random numbers,
they proved a safe way of recycling random numbers,
which drastically promotes efficiency.

In the next reaction method, we store $t_{i}$, the absolute time
when the next event-$i$ occurs, instead of $\tau_{i}$.
In the first step, we have to calculate $t_{i} = \tau_{i}$ for every $i$,
according to the exponential distribution
$P(\tau_{i}) = a_{i} e^{-a_{i}\tau_{i}}$ ($0 < \tau_{i}$).
We choose the event with the smallest $t_{i}$.
According to the event, we set the time $t$ forward to $t_{i}$,
change the state, and recalculate each $a_{i}$.

In steps that follow, we recalculate each $t_{i}$ as follows.
\begin{enumerate}
\item As regards the event just executed, we should recalculate
$t_{i} = t + \tau_{i}$ with the exponential distribution of $\tau_{i}$,
identical to the first step.

\item As regards other events whose frequency $a_{i}$ has changed,
we should recalculate the corresponding $t_{i}$.
For such events, we convert $t_{i}$ as
\[
t_{new} = t_{old} + \frac{a_{old}}{a_{new}} (t_{old} - t).
\]
$t_{old}$ is the $t_{i}$ before the event and
$t_{new}$ is that after the event.

With this conversion, the actual frequency is adjusted
from $a_{old}$ to $a_{new}$, without using random numbers\footnote{Note that Gibson and Bruck mathematically proved the validity of the method
and did not mention numerical errors in their paper.
In some cases, repetition of the conversion might be numerically dangerous.
In case there are various time scales of reactions,
incautious coding may result in numerical errors
(e.g., rare events would not occur forever).}.

\item As for other events whose frequency $a_{i}$ has not changed,
we do not need to recalculate the corresponding $t_{i}$.
\end{enumerate}
Subsequently, we choose the smallest $t_{i}$, and proceed further.

If the event executed does not influence $a_{i}$,
we do not need to recalculate the concerned $a_{i}$ and $t_{i}$ (except for those of
the event just executed).
Thus, it is useful to manage the dependency of $a_{i}$ on each event.
With this intention, we prepare a dependency graph
that shows which $a_{i}$ should be updated after event-$j$
(event-$i$ depends on event-$j$).
In a large reaction network, such as biochemical pathways,
one chemical species can react with only a small part of the chemicals
in the entire system.
In such cases, recalculation is not required for irrelevant chemical species;
hence, we can accelerate simulations with help of a dependency check.

It is also important to find the smallest $t_{i}$ quickly.
For this purpose, we use a heap, a binary tree
in which each node is larger than or equal to its parent.
The root is the smallest at any instance\footnote{The heap sort is an O($N \log N$) sorting algorithm.
If only one $t_{i}$ has changed in the step,
the cost of resorting is O($\log N$).}.

The next reaction method requires only one random number per event,
and executes much faster than first reaction method,
especially in case of many chemicals and reactions\footnote{While the next reaction method is an exact algorithm,
there are also some approximate methods derived from
Gillespie's algorithms, e.g., refs. \cite{Gillespie2001,Arkin2}.}.



\clearpage

\begin{figure}
\begin{center}
\includegraphics[width=69mm]{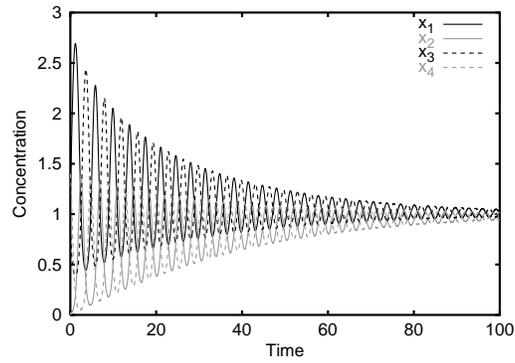}
\end{center}
\caption{Time series of the concentration $x_{i}$ at the continuum limit
for $r=1$, $s=1$, and $D=1/32$.
With $D>0$, the rate equation has a stable fixed point $\forall i : x_{i}=1$.
Each $x_{i}$ converges to the fixed point.}
\label{fig:A-cont}
\end{figure}

\begin{figure}
\begin{center}
\includegraphics[width=69mm]{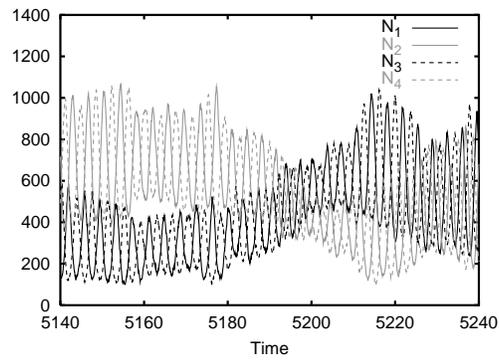}
\end{center}
\caption{Time series of the number of molecules $N_{i}$ for $V=512$ and $D=1/256$.
Each $N_{i}$ fluctuates around the fixed point (as seen in
Fig. \ref{fig:A-cont}) but does not reach $0$.}
\label{fig:A-512}
\end{figure}

\begin{figure}
\begin{center}
\includegraphics[width=69mm]{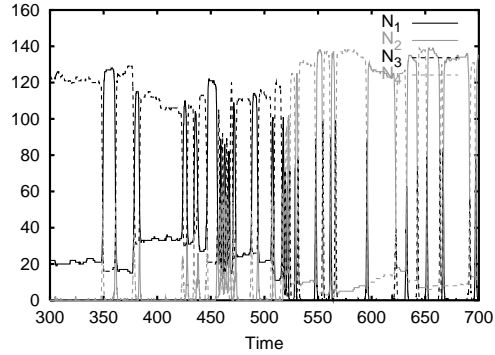}
\end{center}
\caption{Time series of $N_{i}$ for $V=32$ and $D=1/256$.
In this stage, $N_{i}$ can reach $0$,
and the switching states appear.
In the 1-3 rich state, the system successively switches
between the $N_{1} > N_{3}$ and $N_{1} < N_{3}$ states.
The interval of switching is much longer than
the period of continuous vibration ($\approx \pi$)
observed in Figs. \ref{fig:A-cont} and \ref{fig:A-512}.
Around $t=520$, a transition occurs from the 1-3 rich to the 2-4 rich state.}
\label{fig:A-032}
\end{figure}

\begin{figure}
\begin{center}
\includegraphics[width=69mm]{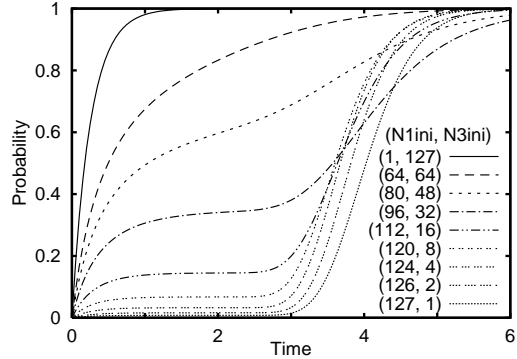}
\end{center}
\caption{Probability for $N_{2}=0$, i.e., reactions have already stalled,
as a function of $t$ (in other words, the cumulative distribution
of the time when $N_{2}$ reaches $0$).
$V=32$, $N_{1 ini} + N_{3 ini} = 128$.
In the case where $N_{1 ini} < N_{3 ini}$ or $N_{1 ini} \approx N_{3 ini}$,
the probability increases just after the reactions start.
For such initial conditions, the reactions terminate near the initial state,
as shown in Fig. \ref{fig:A-swprob-128}.
If $N_{1 ini} \gg N_{3 ini}$, the probability steeply increases at
$t \approx 4$, which corresponds to the switching.
The system takes $\approx 4$ time units to complete the switching.}
\label{fig:A-b0state}
\end{figure}

\begin{thebibliography}{99}


\bibitem{Nicolis} 
 G. Nicolis and I. Prigogine, \textit{Self-Organization in Nonequilibrium Systems}
 (John Wiley, 1977).

\bibitem{Kampen} 
 N. G. van Kampen, \textit{Stochastic processes in physics and chemistry}
 (North-Holland, rev. ed., 1992).

\bibitem{NIO} 
 K. Matsumoto and I. Tsuda,
 ``Noise-Induced Order'',
 Jour. Stat. Phys. \textbf{31}, 87 (1983).

\bibitem{NIP} 
 W. Horsthemke and R. Lefever, \textit{Noise-Induced Transitions},
 edited by H. Haken (Springer, 1984).

\bibitem{SR} 
 K. Wiesenfeld and F. Moss,
 ``Stochastic resonance and the benefits of noise: from ice ages to crayfish and SQUIDs'',
 Nature \textbf{373}, 33 (1995).

\bibitem{TGF1} 
 N. Olsson, E. Piek, P. ten Dijke, and G. Nilsson,
 ``Human mast cell migration in response to members
 of the transforming growth factor-$\beta$ family'',
 Jour. Leukocyte Biol. \textbf{67}, 350 (2000).

\bibitem{IL1B} 
 X. Wang, G. Z. Feuerstein, J. Gu, P. G. Lysko, and T. Yue,
 ``Interleukin-1$\beta$ induces expression of adhesion molecules in human
 vascular smooth muscle cells and enhances adhesion of leukocytes to smooth muscle cells'',
 Atherosclerosis \textbf{115}, 89 (1995).

\bibitem{McAdams} 
 H. H. McAdams and A. Arkin,
 ``It's a noisy business!  Genetic regulation at the nanomolar scale'',
 Trends Genet. \textbf{15}, 65 (1999).

\bibitem{Arkin1} 
 C. V. Rao, D. M. Wolf, and A. P. Arkin,
 ``Control, exploitation and tolerance of intracellular noise'',
 Nature \textbf{420}, 231 (2002).


\bibitem{Elowitz2002} 
 M. B. Elowitz, A. J. Levine, E. D. Siggia, and P. S. Swain,
 ``Stochastic Gene Expression in a Single Cell'',
 Science \textbf{297}, 1183 (2002).

\bibitem{Blumenfeld1991} 
 L. A. Blumenfeld, A. Y. Grosberg, and A. N. Tikhonov,
 ``Fluctuations and mass action law breakdown
 in statistical thermodynamics of small systems'',
 Jour. Chem. Phys. \textbf{95}, 7541 (1991).

\bibitem{Blumenfeld} 
 L. A. Blumenfeld and A. N. Tikhonov,
 \textit{Biophysical Thermodynamics of Intracellular Processes
 --- Molecular Machines of the Living Cell}
 (Springer-Verlag, New York, 1994).



\bibitem{Mikhailov2C} 
 P. Stange, A. S. Mikhailov, and B. Hess,
 ``Coherent Intramolecular Dynamics of Enzymic Reaction Loops in Small Volumes'',
 Jour. Phys. Chem. B \textbf{104}, 1844 (2000).

\bibitem{Mikhailov2002} 
 A. S. Mikhailov and B. Hess,
 ``Self-Organization in Living Cells:
 Networks of Protein Machines and Nonequilibrium Soft Matter'',
 Jour. Biol. Phys. \textbf{28}, 655 (2002).

\bibitem{Togashi} 
 Y. Togashi and K. Kaneko,
 ``Transitions induced by the discreteness of molecules in a small autocatalytic system'',
 Phys. Rev. Lett. \textbf{86}, 2459 (2001).

\bibitem{Togashi2} 
 Y. Togashi and K. Kaneko,
 ``Alteration of Chemical Concentrations through Discreteness-Induced Transitions
 in Small Autocatalytic Systems'',
 Jour. Phys. Soc. Jpn. \textbf{72}, 62 (2003).




\bibitem{KanekoYomoMinority} 
 K. Kaneko and T. Yomo,
 ``On a Kinetic Origin of Heredity:
 Minority Control in a Replicating System with Mutually Catalytic Molecules'',
 Jour. Theor. Biol. \textbf{214}, 563 (2002).

\bibitem{KanekoMinNet2002} 
 K. Kaneko,
 ``Kinetic Origin of Heredity in a Replicating System with a Catalytic Network'',
 Jour. Biol. Phys. \textbf{28}, 781 (2002).

\bibitem{KanekoMinNet2003} 
 K. Kaneko,
 ``Recursiveness, switching, and fluctuations in a replicating catalytic network'',
 Phys. Rev. E \textbf{68}, 031909 (2003).

\bibitem{MinorityExperiment} 
 T. Matsuura, M. Yamaguchi, E. P. Ko-Mitamura, Y. Shima, I. Urabe, and T. Yomo,
 ``Importance of compartment formation for a self-encoding system'',
 Proc. Nat. Acad. Sci. \textbf{99}, 7514 (2002).

\bibitem{Mikhailov1A} 
 B. Hess and A. Mikhailov,
 ``Self-Organization in Living Cells'',
 Science \textbf{264}, 223 (1994).
 
\bibitem{Mikhailov1B} 
 B. Hess and A. Mikhailov,
 ``Microscopic Self-organization in Living Cells: A Study of Time Matching'',
 Jour. Theor. Biol. \textbf{176}, 181 (1995).

\bibitem{Kuramoto1} 
 Y. Kuramoto,
 ``Fluctuations around Steady States in Chemical Kinetics'',
 Prog. Theor. Phys. \textbf{49}, 1782 (1973).

\bibitem{Kuramoto2} 
 Y. Kuramoto,
 ``Effects of Diffusion on the Fluctuations in Open Chemical Systems'',
 Prog. Theor. Phys. \textbf{52}, 711 (1974).

\bibitem{Togashi3} 
 Y. Togashi and K. Kaneko,
 ``Molecular discreteness in reaction-diffusion systems yields steady
 states not seen in the continuum limit'',
 \textit{to appear in} Phys. Rev. E (2004).


\bibitem{Tokita1999} 
 K. Tokita and A. Yasutomi,
 ``Mass extinction in a dynamical system of evolution with variable dimension'',
 Phys. Rev. E \textbf{60}, 842 (1999).

\bibitem{Tokita2003} 
 K. Tokita and A. Yasutomi,
 ``Emergence of a complex and stable network in a model ecosystem with extinction and mutation'',
 Theor. Popul. Biol. \textbf{63}, 131 (2003).


\bibitem{Gillespie1992} 
 D. T. Gillespie,
 ``A rigorous derivation of the chemical master equation'',
 Physica A \textbf{188}, 404 (1992).

\bibitem{MortonFirth1998} 
 C. J. Morton-Firth and D. Bray,
 ``Predicting Temporal Fluctuations in an Intracellular Signalling Pathway'',
 Jour. Theor. Biol., \textbf{192}, 117 (1998).

\bibitem{Gillespie2} 
 D. T. Gillespie,
 ``Exact Stochastic Simulation of Coupled Chemical Reactions'',
 Jour. Phys. Chem. \textbf{81}, 2340 (1977).

\bibitem{Gillespie1} 
 D. T. Gillespie,
 ``General method for numerically simulating stochastic time evolution of coupled chemical-reactions'',
 Jour. Comput. Phys. \textbf{22}, 403 (1976).

\bibitem{Gibson} 
 M. A. Gibson and J. Bruck,
 ``Efficient Exact Stochastic Simulation of Chemical Systems
 with Many Species and Many Channels'',
 Jour. Phys. Chem. A \textbf{104}, 1876 (2000).

\bibitem{Gillespie2001} 
 D. T. Gillespie,
 ``Approximate accelerated stochastic simulation of chemically reacting systems'',
 Jour. Chem. Phys. \textbf{115}, 1716 (2001).

\bibitem{Arkin2} 
 C. V. Rao and A. P. Arkin,
 ``Stochastic chemical kinetics and the quasi-steady-state assumption:
 Application to the Gillespie algorithm'',
 Jour. Chem. Phys. \textbf{118}, 4999 (2003).

\end{thebibliography}
\end{document}